\documentclass[12pt,preprint]{aastex}

\newcommand{\be}{\begin{equation}}
\newcommand{\ee}{\end{equation}}
 
\newcommand{\mearth}{{ M_\earth }} 
\newcommand{\mjup}{{ M_{J} }}

\newcommand{\disturb}{{ {\cal R} }} 
\newcommand{\gamtwo}{{ {\widetilde \gamma} }} 
\newcommand{\atime}{{ \tau_{a} }} 
\newcommand{\etime}{{ \tau_{e} }} 
\newcommand{\ttime}{{ \tau_{T} }} 
\newcommand{\rtime}{{ \tau_{R} }} 

\begin{document} 

\title{\bf EFFECTS OF TURBULENCE, ECCENTRICITY DAMPING, 
AND MIGRATION RATE ON THE CAPTURE OF PLANETS INTO 
MEAN MOTION RESONANCE} 

\author{Jacob A. Ketchum$^1$, Fred C. Adams$^{1,2,3}$, 
and Anthony M. Bloch$^{1,4}$} 

\affil{$^1$Michigan Center for Theoretical Physics \\
Physics Department, University of Michigan, Ann Arbor, MI 48109} 

\affil{$^2$Astronomy Department, University of Michigan, Ann Arbor, MI 48109} 

\affil{$^3$Kavli Institute for Theoretical Physics, University of California, Santa Barbara, 93106} 

\affil{$^4$Department of Mathematics, University of Michigan, 
Ann Arbor, MI 48109} 

\begin{abstract} 

Pairs of migrating extrasolar planets often lock into mean motion
resonance as they drift inward.  This paper studies the convergent
migration of giant planets (driven by a circumstellar disk) and
determines the probability that they are captured into mean motion
resonance.  The probability that such planets enter resonance depends
on the type of resonance, the migration rate, the eccentricity damping
rate, and the amplitude of the turbulent fluctuations. This problem is
studied both through direct integrations of the full 3-body problem,
and via semi-analytic model equations. In general, the probability of
resonance decreases with increasing migration rate, and with
increasing levels of turbulence, but increases with eccentricity
damping. Previous work has shown that the distributions of orbital
elements (eccentricity and semimajor axis) for observed extrasolar
planets can be reproduced by migration models with multiple
planets. However, these results depend on resonance locking, and this
study shows that entry into -- and maintenance of -- mean motion
resonance depends sensitively on migration rate, eccentricity damping,
and turbulence.

\end{abstract} 


\section{INTRODUCTION} 

The past decade has led to tremendous progress in our understanding of
extrasolar planets and the processes involved in planet formation.
These advances involve both observations, which now include the
detection of hundreds of planets outside our Solar System (see, e.g.,
Udry et al. 2007 for a recent review), along with a great deal of
accompanying theoretical work. One surprising result from the
observations is the finding that extrasolar planets display a much
wider range of orbital configurations than was originally anticipated.
Planets thus move (usually inward) from their birth sites, while they
are forming or immediately thereafter, in a process known as planet
migration (e.g., see Papaloizou \& Terquem 2006 and Papaloizou et al.
2007 for recent reviews).

Many of the observed solar systems contain multiple planets, and many
others may be found in the near future. For systems that contain more
than one planet, theoretical work indicates that the migration process
often results in planets entering into mean motion resonance (e.g.,
Lee \& Peale 2002, Nelson \& Papaloizou 2002), at least for some
portion of their migratory phase of evolution. During this epoch,
interacting planets (which are often in or near resonance) tend to
excite the orbital eccentricity of both bodies.  This planet
scattering process, acting in conjunction with inward migration due to
torques from circumstellar disks, can produce broad distributions of
both semi-major axis and eccentricity (Adams \& Laughlin 2003,
Moorhead \& Adams 2005, Chatterjee et al. 2008, Ford \& Rasio 2008);
these distributions of orbital elements are comparable to those of the
current observational sample, although significant uncertainties
remain. In any case, the final orbital elements at the end of the
migration epoch --- and planetary survival --- depend sensitively on
whether or not the planets enter into mean motion resonance.

These systems are highly chaotic, and display extreme sensitivity to
initial conditions, so that the outcomes must be described
statistically. Nonetheless, the distributions of final system
properties are well-defined and depend on whether the planets enter
into mean motion resonance as they migrate inwards; the outcomes also
depend on the type of resonance and how deeply the planets are bound
into a resonant state. The circumstellar disks that drive inward
migration also produce damping and/or excitation (Goldreich \& Sari
2003, Ogilvie \& Lubow 2003) of orbital eccentricity, and this
complication affects the maintenance of resonance. The disks are also
expected to be turbulent, through the magneto-rotational instability
(MRI) and/or other processes (Balbus \& Hawley 1991). With sufficient
amplitude and duty cycle, this turbulence also affects the maintenance
of mean motion resonance (Adams et al. 2008, Lecoanet et al. 2009,
Rein \& Papaloizou 2009), and thereby affects the distributions of
orbital elements resulting from migration (Moorhead 2008).

The goal of this paper is to understand the probability for migrating
planets to enter into mean motion resonance and the probability for
survival of the resulting resonant states. Previous work has shown
that entry into resonance is affected by the migration rate (Quillen
2006), where fast migration acts to compromise resonant states. This
study expands upon previous efforts by considering the effects of not
only the migration rate, but also eccentricity damping and turbulent
forcing on the probability of attaining and maintaining a resonant
state. This paper considers the action of these three variables,
jointly and in isolation, and covers a wide range of parameter space.
In addition, we address the problem through both numerical and
semi-analytic approaches (where `semi-analytic' refers to models where
the equations are reduced to, at most, ordinary differential
equations.)  The results depend on the type of resonance under
consideration; this work considers a range of cases, but focuses on
the 2:1, 5:3, and 3:2 mean motion resonances.

This paper is organized as follows. We first perform a large ensemble
of numerical integrations in Section \ref{sec:numerical}. These
numerical experiments follow two planets undergoing convergent
migration, and include both eccentricity damping and forcing terms due
to turbulent fluctuations.  The results provide an estimate for the
survival of systems in resonance as a function of migration rate,
eccentricity damping rate, and turbulent amplitudes. In order to
isolate the physical processes taking place, we develop a set of model
equations to study the problem in Section \ref{sec:model}. This model,
which follows directly from previous work (Quillen 2006), illustrates
how fast migration rates and high eccentricities act to compromise
resonance. The paper concludes, in Section \ref{sec:conclude}, with a
summary of results and a discussion of their implications for observed
extrasolar planets.  Finally, we present a phase space analysis of the
model problem in an Appendix. 

\section{NUMERICAL INTEGRATIONS} 
\label{sec:numerical} 

\subsection{Formulation} 

In this section we consider the direct numerical integration of
migrating planetary systems, i.e., we integrate the full set of 18
phase space variables for the 3-body problem consisting of two
migrating planets orbiting a central star.  For most of our
simulations, the planets are started in the same plane so that the
dynamics is only two dimensional; however, we have also run cases that
explore all three spatial dimensions.  The integrations are carried
out using a Burlisch-Stoer (B-S) integration scheme (e.g., Press et
al.  1992). In addition to gravity, we include forcing terms that
represent inward migration, eccentricity damping, and turbulence. All
three of these additional effects arise to the forces exerted on the
planet(s) by a circumstellar disk. In this context, however, we do not
model the disk directly, but rather include forcing terms to model its
behavior, as described below.

To account for planet migration, we assume that the semimajor axis of
the outer planet decreases with time according to the ansatz
\be
{1 \over a} {da \over dt} = - {1 \over \atime} \, , 
\label{adamp} 
\ee 
where $\atime$ is the migration time scale.  Further, we assume that
only the outer planet experiences torques from the circumstellar disk.

Small planets, those with masses smaller than that of Saturn, cannot
clear gaps in their circumstellar disks and tend to migrate inward
quickly in a process known as Type I migration (e.g., Ward 1997ab). A 
number of studies have shown that the Type I migration rate depends on
the disk thermal properties and on local gradients of the gas density
(e.g., Baruteau \& Masset 2008, Paardekooper \& Papaloizou 2008,
Masset \& Casoli 2009, Paardekooper et al. 2010). As a result, for
some disks, Type I migration can be much slower (sometimes even
directed outward) and a wide range of migration rates is possible. 
Larger bodies clear gaps and migrate more slowly. Estimates of the
migration timescale for planets with $a \sim 1$ AU typically fall in
the range $10^4 - 10^6$ yr (e.g., see Goldreich \& Tremaine 1980,
Papaloizou \& Larwood 2000).  

Planets are thus expected to experience a range of migration rates,
depending on both planet masses and disk properties. Since we want to
isolate the effects of migration rate on entry into resonance, we
adopt a purely parametric approach.  We thus consider a wide range of
migration rates, where the migration timescale varies over the range
$\atime$ = $10^3 - 10^6$ yr. Note that the shorter time scales are
included here to study the physics of resonance capture (at these fast
rates) and are not generally expected in most circumstellar disks. On
the other hand, fast migration could occur for planets with mass
$\sim10 \mearth$ migrating within circumstellar disks that have
sufficiently small aspect ratios ($H/r < 0.03$) and large masses
(Masset \& Papaloizou 2003).

In addition to inward migration, circumstellar disks also tend to damp
orbital eccentricity $e$ of the migrating planet. This damping is
generally found in numerical simulations of the process (e.g., Lee \& 
Peale 2002, Kley et al. 2004), and can be parameterized such that
\be
{1 \over e} {de \over dt} = - {1 \over \etime} = 
K \left( {1 \over a} {da \over dt} \right) 
\qquad {\rm so} \, \, {\rm that} \qquad \etime = \atime / K \, , 
\label{edamp} 
\ee
where $\etime$ is the eccentricity damping timescale. Some analytic
calculations suggest that eccentricity can be excited through the
action of disk torques (Goldreich \& Sari 2003, Ogilvie \& Lubow
2003), although multiple planet systems would be compromised if this
were always the case (Moorhead \& Adams 2005). Additional calculations
show that disks generally lead to both eccentricity damping and
excitation, depending on the disk properties, gap shapes, and other
variables (e.g., Moorhead \& Adams 2008).  The value of the damping
parameter $K$ can also be inferred from hydrodynamical simulations,
which predict a range of values. Studies of resonant systems (Kley et
al. 2004) advocate $K$ values of order unity.  In isothermal disk
models, however, $K \sim 10$ for typical cases (e.g., Cresswell \&
Nelson 2008). More recent work indicates that in radiative disk
models, the eccentricity damping parameter $K$ can be as large as 100
(Bitsch \& Kley 2010).

In spite of the aforementioned uncertainties, this study focuses on
the case of pure damping, adopts fixed values of $K$ for a given
simulation, and considers its effects on the dynamics of mean motion
resonances.  We expect that the inclusion of damping will act to
enhance the survival of mean motion resonances (Lecoanet et al.
2009). Using the ansatz of equation (\ref{edamp}), this study
considers a wide range of the damping parameter $K$ such that 
$0 \le K \le 100$, where we consider the cases $K$ = 1 and $K$ = 10 as
our ``standard'' values.

Turbulence is included by applying discrete velocity perturbations at
regular time intervals; for the sake of definiteness, the forcing
intervals are chosen to be twice the orbital period of the outer
planet (four times the period of the inner planet for systems with 2:1
period ratio).  Both components of velocity in the plane of the orbit
are perturbed randomly, but the vertical component of velocity is not
changed. The amplitude of the velocity perturbations thus represents
one of the variables that characterize the system. These amplitudes
are chosen to be consistent with the expected torques, as described
below.

The torques due to turbulent fluctuations have been studied previously
using MHD simulations (e.g., Nelson \& Papaloizou 2004; Laughlin et al.  
2004; Nelson 2005; Oishi et al. 2007), and these results can be used
to estimate the range of amplitudes. The torque exerted on a planet by
a circumstellar disk can be expressed as a fraction of the benchmark
torque $T_D$ = $2 \pi G \Sigma r m_P$, where $\Sigma$ is the surface
density of the disk, $r$ is the radial coordinate, and $m_P$ is the
planet mass (Johnson et al. 2006). The scale $T_D$ thus serves as a
maximum torque in this problem. The amplitude of the expected angular
momentum fluctuations is then given by $\Delta J$ = $f_T \Gamma_R T_D
t_T$, where $f_T$ is the fraction of torque scale $T_D$ realized by
the disk, $\Gamma_R$ is a reduction factor due to planets creating
gaps in the disk, and $t_T$ is the time required for the disk to
produce an independent realization of the turbulence.  Previous work
suggests that $f_T \sim 0.05$ (Nelson 2005), $\Gamma_R \sim 0.1$
(Adams et al. 2008), and $t_T$ is comparable to the orbit time of the
outer planet (Laughlin et al. 2004, Nelson 2005).  Including all of
these factors, we expect that $[(\Delta J)/J] \sim 10^{-4}$ under
typical conditions (a disk mass of $\sim 0.05$ $M_\odot$ with
well-developed MRI turbulence such that $\alpha \sim 10^{-3}$). Under
some circumstances, the equatorial plane of the disk is not
sufficiently ionized to support MRI turbulence and a dead zone
develops; in this case, the fraction $f_T$ would be dramatically
decreased.  Given the uncertainties in turbulent behavior, and the
wide range of possible disk conditions, the fluctuation amplitude
could vary by an order of magnitude (perhaps more) in either
direction. As a result, we consider turbulent fluctuation amplitudes
in the range $0 \le [(\Delta J)/J] \le 10^{-3}$.

For a given realization of the migration scenario, we need to
determine whether or not the system resides in mean motion resonance.
First, we determine the ratio of the orbital periods of the two planets. 
It is straightforward to determine when systems have nearly integer
period ratios and this condition can be used as a proxy for being in a
mean motion resonance. However, this condition is necessary but not
sufficient, so we must also monitor the relevant resonance angles
(Murray \& Dermott 1999; hereafter MD99).  For first order resonances,
these angles have the form
\be 
\theta_1 = (j + 1) \lambda_2 - j \lambda_1 - \varpi_1 \, , 
\label{firstangle} 
\ee 
\be 
\theta_2 = (j + 1) \lambda_2 - j \lambda_1 - \varpi_2 \, , 
\ee 
\be 
\theta_3 = \varpi_1 - \varpi_2 \, .  
\ee 
For second order resonances, these angles take the form 
\be 
\theta_1 = (j + 2) \lambda_2 - j \lambda_1 - 2 \varpi_1 \, , 
\ee 
\be 
\theta_2 = (j + 2) \lambda_2 - j \lambda_1 - \varpi_1 - \varpi_2 \, , 
\ee 
\be 
\theta_3 = (j + 2) \lambda_2 - j \lambda_1 - 2 \varpi_2 \, , 
\ee 
\be
\theta_4 = \varpi_1 - \varpi_2 \, .  
\label{lastangle} 
\ee 
In order to monitor the angles, and determine if the system is in a
resonant state, we must choose the appropriate time windows.  Note
that the resonance angle $\theta_0 = \varpi_1 - \varpi_2$ oscillates
on a much longer timescale than the other angles (where $\theta_0$ =
$\theta_3$ ($\theta_4$) for first (second) order resonances).  As a
result, the angle $\theta_0$ is measured over a time period
corresponding to 1500 orbits of the outer planet, whereas the other
angles (which oscillate faster) are monitored over a time window of
300 orbits of the outer planet. These timescales are chosen to be
(roughly) several times the expected libration periods of the angles
(and the expected libration periods can be calculated from the
restricted three-body problem --- see MD99).  Each angle is considered
to be in libration if its value stays bounded within 120 degrees of
the effective stability point for the time periods given above. In
this context, the effective stability point is determined by the mean
value of the angle over the given time window for monitoring; note
that these systems are highly interactive (e.g., due to turbulent
forcing) so that the stability points are not fixed. Notice also that
the value of 120 degrees was chosen arbitrarily. If any of the angles
$\theta_i$ obtain a value greater than 120 degrees, measured from
either side of the effective stability point, then that angle is
considered to not be in resonance.  The code continues to monitor all
of the relevant angles for the duration of the time when the periods
have a well-defined ratio (of small integers).  As a result, each
resonance angle could go in and out of libration many times.

\subsection{Numerical Results for Resonance Survival} 

Given the formulation described above, we numerically study the entry
of planets into mean motion resonance and the subsequent survival of
the resonant configurations. These results depend on a number of
parameters, including the migration rate, the eccentricity damping
rate, the level of turbulence, and the planetary masses. As indicated
by the semi-analytic models (Section \ref{sec:model}, Quillen 2006),
we expect the survival of mean motion resonance to be compromised with
sufficiently fast migration rates. The introduction of turbulence can
act to further reduce the ability of systems to stay in resonance
(Adams et al. 2008), whereas eccentricity damping generally acts in
the opposite direction by helping to maintain resonance (Lecoanet et
al. 2009). The results also depend on the masses of the planets. As
the masses increase, the systems become more highly interactive and
mean motion resonance is harder to maintain.

For the first set of simulations, we begin with a standard two planet
system consisting of a Jovian mass planet $m_1 = 1 \mjup$ and a
``super-Earth'' with the mass $m_2$ = 10 $\mearth$.  The properties of
this system are close to the restricted 3-body problem and hence the
resonance is expected to be described reasonably well by the pendulum
model of MD99.  The star is taken to have a mass $M_\ast$ = 1.0
$M_\odot$. The Jovian planet acts as the inner planet and begins with
an eccentricity $e_1$ = 0.05 and a period of $P_1$ = 1000 days (so
that $a_1$ = 1.96 AU).  The smaller planet starts with an eccentricity
$e_1$ = 0.10 and a semi-major axis of $a_2$ = 1.8 $a_1$, equivalent to
a period ratio of $P_2/P_1$ = 2.4, which places the system comfortably
outside the 2:1 mean motion resonance. Both of the planets are placed
in the same orbital plane.  As the outer planet migrates inward, it
can (in principle) enter into the 2:1 resonance; if the migrating
planet passes through the 2:1 resonance, it can then (potentially)
enter into resonant states with smaller period ratios. 

In this parametric study, we allow migration of the outer planet,
given by equation (\ref{adamp}), to continue throughout the
simulations.  Since the inner Jovian planet is expected to open a gap
in the circumstellar disk however, the migration rate could be 
altered, where the variations depend on the gap structure. Although
not considered herein, some disks with gaps can even halt migration
altogether and produce planet traps (Masset et al. 2006). In addition,
since the (smaller) outer planet often acquires substantial
eccentricity, it will move in and out of the gap over the course of
its orbit. This effect leads to time dependent migration torques that
vary on the orbital timescale; the time variations tend to average out
over the libration timescale of the resonances, but the migration rate
could be altered slightly.

Because these systems are highly chaotic, different realizations of
the problem lead to different outcomes. For each set of parameters, 
we perform an ensemble of (at least) 1000 effectively equivalent
simulations, where the simulations differ only by the relative
position of the two planets in their orbits and by the relative angle
between the orientation of the two orbits (i.e., the arguments of
periastron $\varpi_2 - \varpi_1$).  The length of the numerical
integration $t_T$ is set by the migration timescale $\atime$, such
that $t_T$ = $\atime$ for the slowest migration rate ($\atime$ =
$10^6$ yr) and $t_T \approx 10 \atime$ for the fastest migration rate
($\atime$ = $10^3$ yr). The overall integration times are thus shorter
for the faster migration rates, but remain long enough to emcompass
many libration time scales for the relevant resonance angles.

\begin{figure} 
\figurenum{1} 
{\centerline{\epsscale{0.90} \plotone{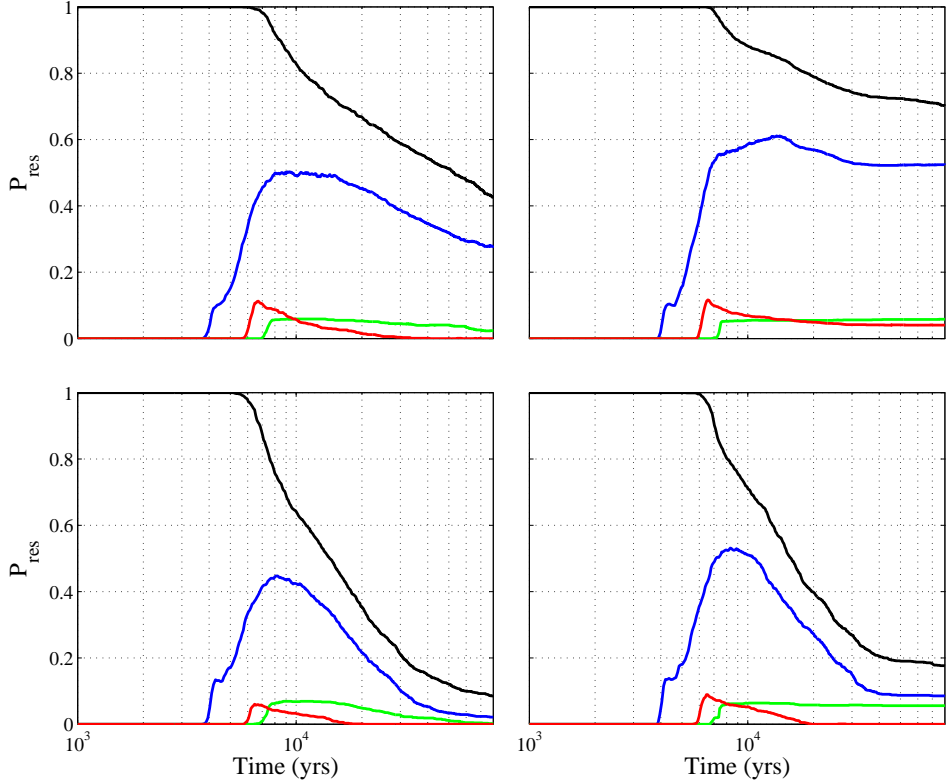} } } 
\figcaption{Fraction of systems in resonance as a function of time.
Each curve shows the fraction of the ensemble that reside in 2:1
resonance (blue curve), 5:3 resonance (red curve), and 3:2 resonance
(green curve) versus time. The systems are considered to be in
resonance if the period ratios are near the relevant integer values
and any of the resonance angles are librating (see text).  The black
curves shows the fraction of systems that remain intact, without
losing a planet, as a function of time.  For the cases shown here, the
migration timescale $\atime$ = $2 \times 10^4$ yr. The two panels on
the top include eccentricity damping with parameter $K$ = 1; the two
panels on the left include turbulence with the standard level of
fluctuations (see text). } 
\label{fig:tevolve} 
\end{figure}

\begin{figure} 
\figurenum{2} 
{\centerline{\epsscale{0.90} \plotone{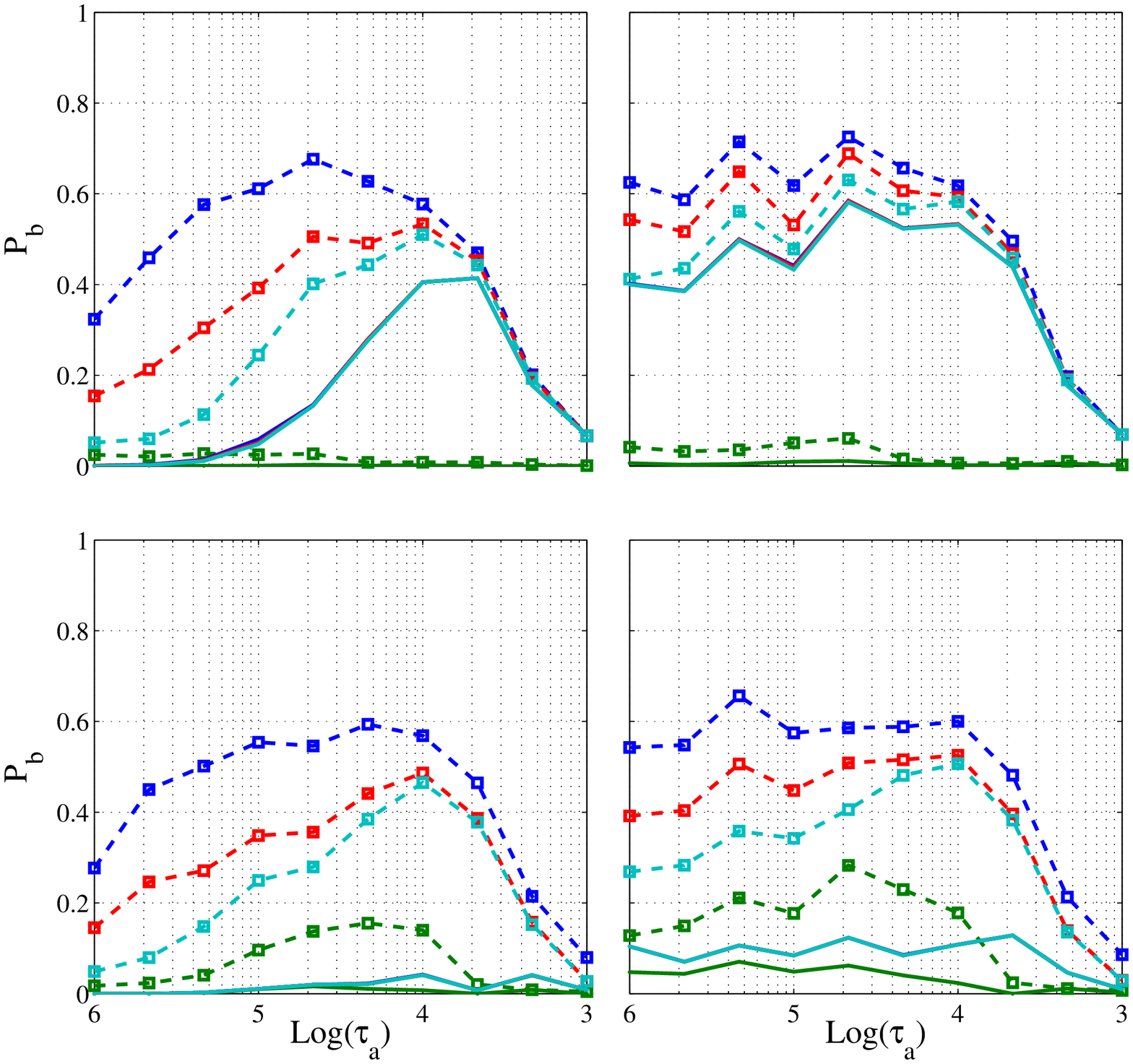} } } 
\figcaption{Fraction of systems in 2:1 resonance according to four
criteria: The curves show the fraction of the ensemble that have
nearly 2:1 period ratio (blue dashed curve), are in $\theta_1$
resonance (cyan dashed curve), are in $\theta_2$ resonance (red dashed
curve), and are in $\theta_3$ resonance (green dashed curve). The
heavy solid curves show the fraction of the ensemble that are in
resonance (of each type) at the end of the migration epoch. Note that
the fractions of $\theta_1$ and $\theta_2$ resonances at the end of
the epoch are nearly identical. }
\label{fig:criteria} 
\end{figure}

\subsubsection{Time Evolution and Resonance Criteria} 

For this set of system parameters, Figure \ref{fig:tevolve}
illustrates the basic time evolution for an ensemble of planetary
systems. Here, the fractions of the systems that reside in mean motion
resonance are plotted as a function of time for a moderately short
migration timescale $\atime$ = $2 \times 10^4$ yr. The simulations
shown in the top panels include eccentricity damping with parameter
$K$ = 1; the two panels on the left include turbulence with the
standard level of fluctuations $(\Delta J)/J \sim 2 \times 10^{-4}$. 
The various curves in each panel correspond to resonances with period
ratios of 2:1 (blue), 5:3 (red), and 3:2 (green).  The systems are
considered to be in resonance if the period ratios are near the
relevant integer values and if any of the resonance angles are
librating (see Section 2.1 and equations [\ref{firstangle} --
  \ref{lastangle}]).  Note that this migration rate was used because
slower migration rates lead to few systems in the 5:3 and 3:2
resonances.  Since all of the systems start outside of resonance, the
fractions start at zero and increase with time as migration pushes the
planets together. The 2:1 resonance is encountered first, so that
corresponding fraction grows first. As the systems evolve, resonance
is often compromised, so that the fractions reach a peak value and
then decrease.  After some of the systems leave the 2:1 state, they
become locked into the 5:3 resonance, and then sometimes the 3:2
resonance. As a result, the peak fraction occurs later for resonances
that are further inward, and the peak is lower for the weaker
resonances (as expected). When systems begin to leave resonance, some
of them decay by losing a planet through ejection, accretion onto the
star, or collision with the other planet (the probabilities of these
end states are quantified in Section 2.3). This effect is illustrated
in Figure \ref{fig:tevolve} by the black curves, which show the
fraction of systems that retain both planets as a function of time.

We note that planetary systems are often said to ``be in resonance''
according to different criteria. This trend is illustrated in Figure
\ref{fig:criteria} for the case of the 2:1 mean motion resonance. As
shown in Figure \ref{fig:tevolve}, the fraction of systems that reside
in resonance is a function of time for a given migration rate.  The
peak value of this time-dependent curve can be used as one measure of
the fraction of systems that are in resonance.  However, systems enter
and leave resonance at different times, so that the total fraction of
systems that enter resonance will be larger than the maximum fraction
that reside in resonance at a given time (the peak of this curve).  A
necessary (but not sufficient) condition for a system to be bound into
resonance is for ratio of the periods to be near 2:1. In this context,
the period ratio is ``near'' 2:1 if $|P_2/P_1 - 2| \le 0.01$, where we
discuss this issue more quantitatively below.  As outlined above, we
first invoke the constraint that $P_2/P_1 \approx 2$. This fraction is
shown as the dashed blue curves marked by squares in Figure
\ref{fig:criteria}. The four panels show the effects of including
eccentricity damping (with $K$ = 1, panels in top row) and turbulent
forcing (with $(\Delta J)/J \sim 2 \times 10^{-4}$, panels on left
side).  Next we note that each of the resonance angles can be either
librating or circulating.  For those that are librating, the range of
angles (the libration width) is highly variable. For this paper we use
the requirement that the resonance angles are confined to be within
120 degrees of the effective stability point (as defined above).  With
this specification, the corresponding fractions of systems in
resonance are shown as the cyan curves ($\theta_1$ angle), the red
curves ($\theta_2$ angle), and the green curves ($\theta_3$ angle).
The solid curves in each panel show the fraction of systems for which
any of the resonance angles are librating at the end of the migration
epoch. Note that only a relatively small fraction of the systems
maintain resonance for the entire migration epoch. In addition, the
inclusion of eccentricity damping (top panels) is crucial for the
survival of resonant states.

For completeness, we note that the curves shown in Figure
\ref{fig:criteria} have slightly different meanings for the different
resonance angles. In order for any one of the angles to be considered
in resonance, it must librate over (approximately) three libration
periods.  However, these periods are not the same for the three
angles.  In particular, the libration period for $\theta_3$ is much
longer than the other two.  In addition, for this class of systems,
the orbit of the outer (lighter) planet varies much more than that of
inner Jovian planet. As a result, the argument of periastron of the
outer planet can circulate on a long timescale, but the resonance
angle $\theta_2$ can still be considered (according to the criteria
used here) to be librating.

In order to understand how the period ratios vary, we monitor the
period ratio for systems that are found in 2:1 mean motion resonance.
Monitoring is triggered by the condition that $P_2/P_1 < 2.05$;
however, once triggered, this bound is relaxed and the period ratio
for systems in resonance can take any value as long as the angles are
librating (see above).  We find that systems typically exhibit both a
slight offset from exact commensurability and variations about this
offset.  The offset is typically less than $\sim1\%$ and the standard
deviation is $\sim2\%$.  We note that offsets and variations of this
magnitude are expected, given the size of the terms in the disturbing
function, and hence the size of the non-Keplerian velocities due to
resonance.  Both the offset value and amount of variation depend on
the levels of damping and turbulence, and on the duration of
resonance.  In the absence of turbulence, we find an offset such that
$P_2/P_1 \sim 2.008$. For systems that do not include eccentricity
damping, the variation of the period ratio $\sigma \sim 0.04$, but
decreases for systems that stay in resonance over long times (10 times
the migration time scale $\atime$).  For cases with eccentricity
damping parameter $K$ = 10, the resonances are longer lasting, and we
find $\sigma \sim 0.015$. For systems that include turbulent forcing,
the period ratio $P_2/P_1 \sim 2.007$ with $\sigma \sim 0.07$ for
short lived resonances, but decreases to $P_2/P_1 \sim 2.002$ with
$\sigma \sim 0.015$ for longer lived resonances.

\subsubsection{Resonance Survival} 

\begin{figure} 
\figurenum{3} 
{\centerline{\epsscale{0.90} \plotone{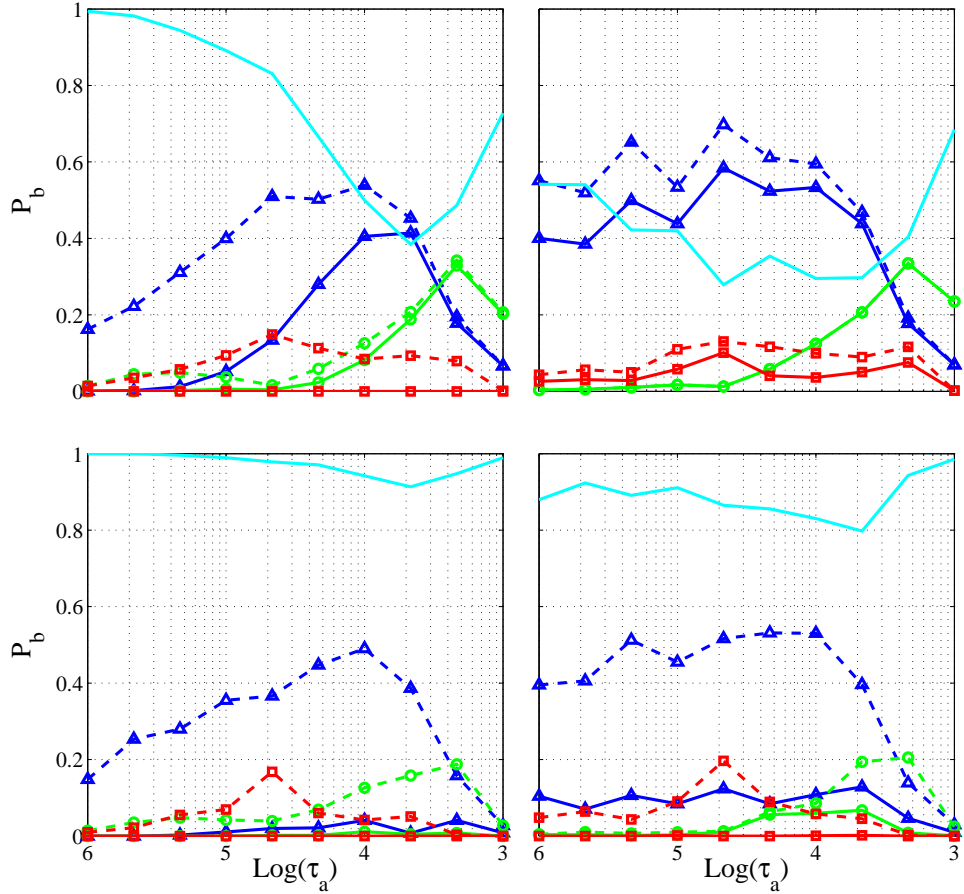} } } 
\figcaption{Effects of eccentricity damping and turbulent forcing on 
the survival of mean motion resonances for a two planet system. 
The inner planet has mass $m_1 = 1 \mjup$, and the outer planet has
mass $m_2 = 10 \mearth$. The two panels in the left column include
turbulent forcing. The panels in the top row include eccentricity
damping, which acts on the same timescale as the migration rate
(eccentricity damping parameter $K$ = 1). The lower right panel shows
the results with migration only. The dashed curves show the fraction
of systems that enter into mean motion resonance as a function of
migration rate ($\atime$ measured in yr), where this fraction is
measured using the peak value (as a function of time -- see the curves
of Figure \ref{fig:tevolve}).  The solid curves show the fraction that
remain in resonance at the end of the migration epoch. The colors
denote the various resonances, including the blue curve marked by
triangles (2:1), red curve marked by squares (5:3), and green curve
marked by circles (3:2). The upper cyan curve shows the fraction of
systems that are not in resonance at the end of the simulations. }
\label{fig:basicrat} 
\end{figure}

Figure \ref{fig:basicrat} shows the effects of both turbulence and
eccentricity damping on the survival of resonances as a function of
migration rate. In this case, we define resonance using the
requirement that the planets have nearly integer period ratios and the
libration width is less than 120 degrees for any of the resonant
angles.  The lower right panel shows the survival of resonances as a
function of migration rate with no eccentricity damping and no
turbulence. This case is thus analogous to the model equations derived
in Section \ref{sec:model} below. As expected, systems tend to leave
resonance if the migration rate becomes too large. Here, systems leave
the 2:1 resonance (blue curve) when the migration rate exceeds roughly
$2 \times 10^{-4}$ yr$^{-1}$ ($\atime$ = 5000 yr).  After leaving the
2:1 resonances, systems can become locked into the 5:3 resonance (red
curve), and/or the 3:2 resonance (green curve). Note that the curve
for 2:1 resonances shows a broad maximum near the migration rate of
$10^{-5}$ yr$^{-1}$ ($\atime$ = 0.1 Myr), with decreasing probability
towards both slower and faster rates. The decrease with increasing
migration rate is expected. The decrease toward slower migration rates
occurs because some of the systems are locked into higher order
resonances, which include the 7:3 and the 9:4 mean motion resonances
(these fractions are not shown in the figure).  With the starting
period ratio of 2.4, the systems must pass through these states to
reach the 2:1 resonance; with extremely slow migration rates, these
weak resonances can (sometimes) survive and thus reduce the
probability of the systems entering the 2:1 resonance.

The effects of including turbulent fluctuations are shown by the
analogous curves in the lower left panel. Turbulence only has a chance
to act on long timescales, so that the simulations with long migration
times (low migration rates) are affected the most. More specifically,
for migration rates slower than about $10^{-5}$ yr$^{-1}$ ($\atime$ =
0.1 Myr), turbulence has time to act, and the probability of
maintaining a resonant configuration is lower, as shown on the left
hand side of the plot.  We note that with the inclusion of turbulence,
the weak higher order resonances (7:3 and 9:4) generally do not
survive (unlike the case of no turbulence in the lower right panel).

The effects of including eccentricity damping is shown by the top
right panel, where we have taken $K$ = 1 (so that eccentricity is
damped on the same timescale that migration takes place; see equation
[\ref{edamp}]). The inclusion of this damping effect acts to preserve
resonance -- note that all of the survival fractions are higher when
${\dot e} \ne 0$ than in the absence of damping. This effect is
especially important for the long-term survival of the resonant states
(compare the solid curves in the top panels with those in the bottom
panels), especially for the case of the 2:1 resonance.  The survival
probabilities of the (weaker) 5:3 and 3:2 resonances are also enhanced
by the inclusion of eccentricity damping, but the absolute values of
these probabilities remain low. Keep in mind that these results
correspond to $K$ = 1; larger eccentricity damping rates lead to more
dramatic consequences (see below).

When both turbulent forcing and eccentricity damping are included, we
obtain the results shown in the upper left panel of Figure
\ref{fig:basicrat}. In this case, the effects of turbulence dominate
at low migration rates, so that fewer resonant systems survive. At
high migration rates, however, turbulence does not have sufficient
time to act and the effects of eccentricity damping lead to a net gain
in the survival fractions. For migration rates faster than about 
$3 \times 10^{-5}$ yr$^{-1}$ ($\atime \approx 0.033$ Myr),
eccentricity damping dominates over the effects of turbulence, so that
more resonant systems survive.

\begin{figure} 
\figurenum{4} 
{\centerline{\epsscale{0.90} \plotone{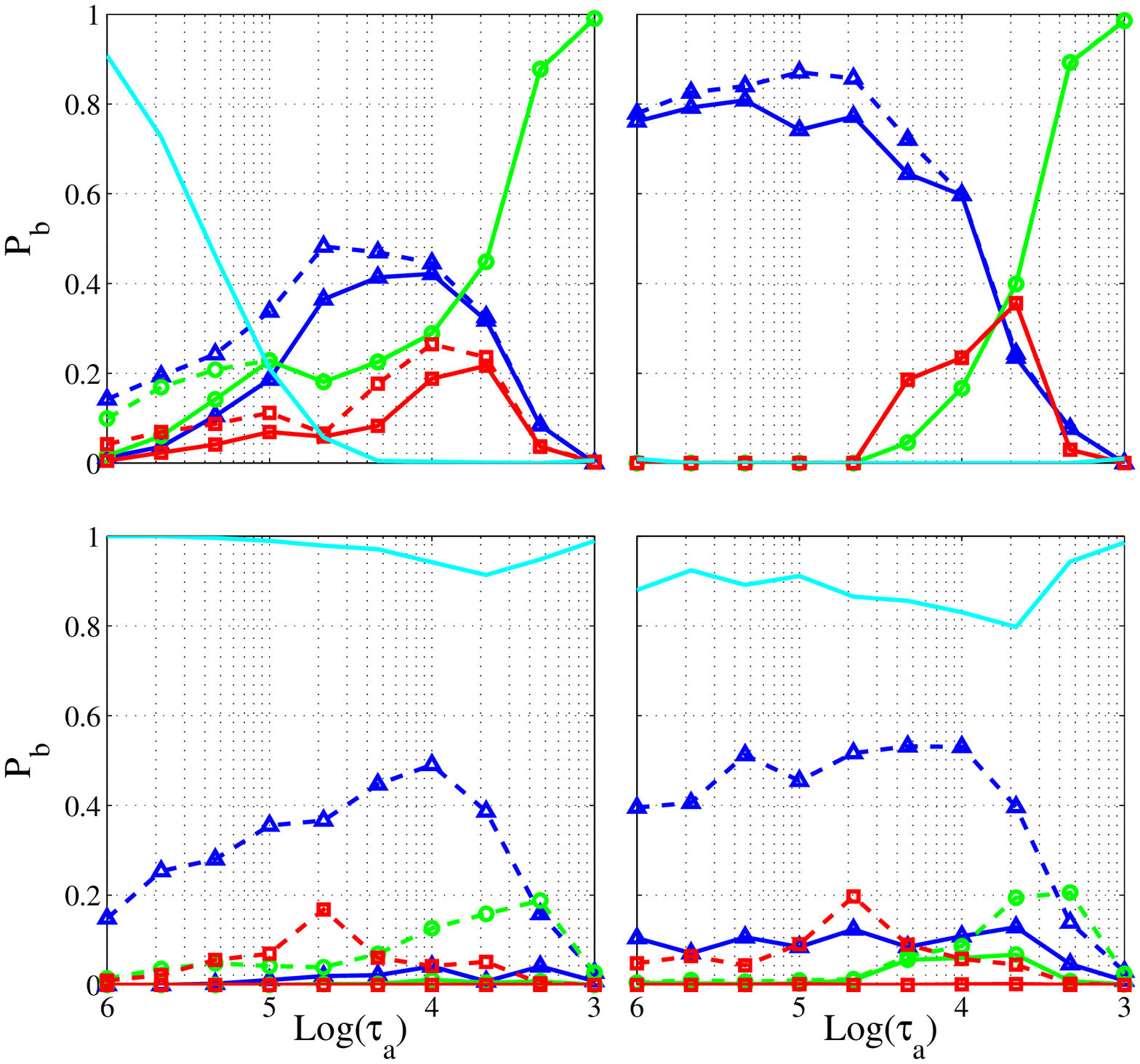} } } 
\figcaption{Effects of eccentricity damping and turbulent forcing on 
the survival of mean motion resonances for a two planet system with 
eccentricity damping parameter $K$ = 10. Other properties are the same
as in Figure \ref{fig:basicrat}: The planet masses are $m_1 = 1 \mjup$
and $m_2 = 10 \mearth$. The panels in the left column include
turbulent forcing. The panels in the top row include eccentricity
damping with $K$ = 10.  The lower right panel shows the results with
migration only. The dashed curves show the fraction of systems that
enter into mean motion resonance as a function of migration rate. The
solid curves show the fraction that remain in resonance at the end of
the migration epoch. The colors denote the various resonances,
including the blue curve marked by triangles (2:1), red curve marked
by squares (5:3), and green curve marked by circles (3:2). The upper
cyan curve shows the fraction of systems that are not in resonance at
the end of the simulations. }
\label{fig:basicten} 
\end{figure}

For comparison, we consider the survival of resonant systems for the
case with eccentricity damping parameter $K$ = 10. These results are
shown in Figure \ref{fig:basicten}, where all of the system parameters
are the same as in Figure \ref{fig:basicrat} except for the larger
rate of eccentricity damping. As expected (e.g., Lecoanet et al.
2009), the simulations with $K$ = 10 result in a larger survival rate
than the corresponding cases with $K$ = 1 (compare the upper right
panels of Figures \ref{fig:basicten} and \ref{fig:basicrat}). For
slower migration rates, where the dominant outcome is the 2:1
resonance (blue curves), the survival rate increases only modestly,
from $P_b \sim 0.6$ to $P_b \sim 0.8$ with increasing values of $K$
(for the case with no turbulence). For higher migration rates, the 3:2
resonance is most common state, and the survival rate increases
substantially for the $K$ = 10 case (compared to $K$ = 1 systems).
For the simulations that include turbulent fluctuations, however, the
differences in survival fractions for the 2:1 resonance between the
$K$ = 10 and $K$ = 1 cases are minimal (compare the upper left panels
of Figures \ref{fig:basicrat} and \ref{fig:basicten}). In the absence
of turbulence, the increase in resonance survival (for larger $K$)
arises most strongly at slow migration rates; however, the regime of
slow migration is where turbulence has enough time to act, and hence
to compromise resonant states. For 3:2 resonances, which arise 
primarily at fast migration rates where turbulence does not have 
enough time to act, the increased eccentricity damping rate leads 
to substantially larger survival fractions. 

\begin{figure} 
\figurenum{5} 
{\centerline{\epsscale{0.90} \plotone{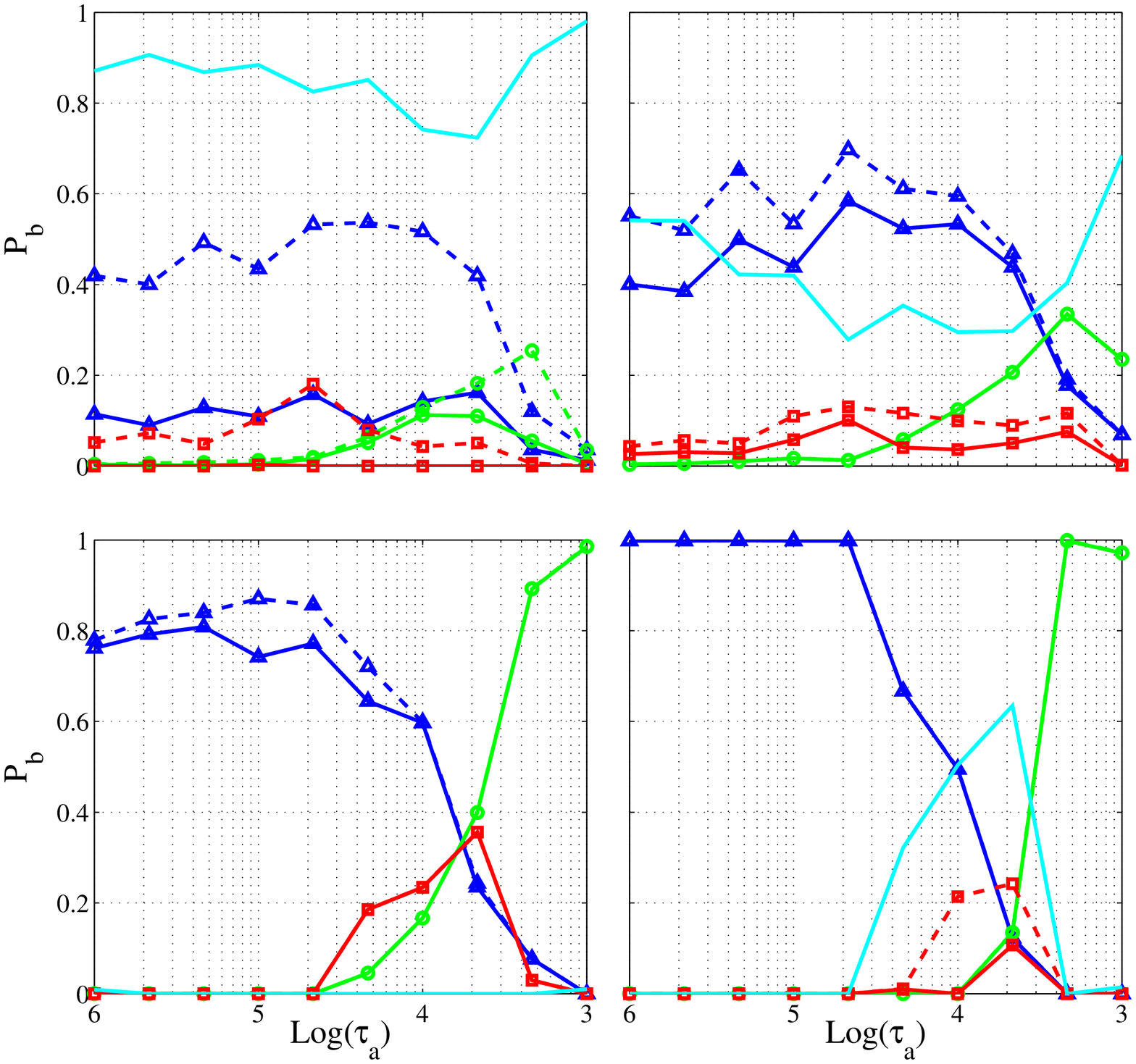} } } 
\figcaption{Effects of eccentricity damping on the survival of mean 
motion resonance. This two-planet system has planetary masses of 
$m_1 = 1 \mjup$ and $m_2 = 10 \mearth$. The four panels show the
survival fractions as a function of migration rate for increasing
values of the eccentricity damping parameter $K$ where ${\dot e}/e$ =
$K {\dot a}/a$. Results are shown for $K$ = 0.1, 1, 10, and 100, where
the $K$ values increase from upper left to lower right. In each panel,
the curves correspond to various resonances, including the blue curve
marked by triangles (2:1), red curve marked by squares (5:3), and 
green curve marked by circles (3:2).  The unmarked cyan curve shows 
the fraction of systems that are not found in any of the mean motion
resonances. The solid curves show the fraction of systems in
resonance at the end of the migration epoch; the dashed curves show
the largest value of the fractions during the migration epoch. }
\label{fig:edamp} 
\end{figure}

Before leaving this section, we note that the simulations shown thus
far all start with the planets confined to the orbital plane. In order
to see how nonzero inclination angles affect the results, we have
carried out a series of test simulations where the planets are given
non-zero displacements in the vertical dimension in their starting
states (so that these simulations are fully three-dimensional). The
results of these test simulations indicate that the third dimension is
unimportant as long as the initial departures from the plane are not
too large.  More specifically, the starting vertical coordinates $z_0$
are uniformly sampled within the range $[-H, H]$, where $H$ is the
scale height of the disk. These test simulations use a variety of
scale heights, with $H/r$ = 0, 0.05, 0.10, and 0.20; we also use a 10
$\mearth$ outer planet, eccentricity damping parameter $K$ = 10, and
our standard level of turbulence. For this choice of parameters, the
results are virtually unchanged.

\subsubsection{Effects of Eccentricity Damping and Turbulence} 

Next we consider the case of eccentricity damping acting alone. Figure
\ref{fig:edamp} shows the survival probabilities as a function of
migration rate for four values of the eccentricity damping timescale,
where the parameter $K$ = 0.1, 1, 10, and 100.  Taken together, the
four panels of Figure \ref{fig:edamp} show that eccentricity damping
acts to increase the fraction of systems that remain in mean motion
resonance. The effect is most pronounced for the 2:1 resonance, and
for slow migration rates. In the regime of slow migration, a
significant fraction of the systems leave the 2:1 resonance,
presumably through the excitation of eccentricity via planet-planet
interactions (Adams \& Laughlin 2003, Moorhead \& Adams 2005,
Chatterjee et al. 2008, Ford \& Rasio 2008, Matsumura et al. 2010).
The inclusion of eccentricity damping counteracts this excitation and
allows more systems to remain in resonance.

For sufficiently large eccentricity damping rates (characterized by
$K$ = 100), essentially all systems remain in 2:1 resonance until the
migration rate exceeds a well-defined value, found numerically to be
$|{\dot a}|/a \sim 3 \times 10^{-5}$ yr$^{-1}$ or $\atime \approx
0.033$ Myr (shown in the lower right panel of Figure \ref{fig:edamp}).
These results show that the loss of resonant states for the other
cases (5:3 and 3:2) also occurs at well-defined values of the
migration rate. In addition, as a rough approximation, the migration
rates at which these three resonances are compromised are found to be
evenly spaced logarithmically (by factors of $\sim3$). This behavior
can be understood in a qualitative manner through simple physical
considerations (see below) and through model equations (Section
\ref{sec:model}). Finally, we note that the 5:3 resonance, which is
second order and hence weak, is sparsely populated; as a result, many
of the systems with migration time scales $\sim3000$ yr are not found
in any resonance.

The basic clock that determines the dynamics of these planetary
systems is set by the libration timescale of the resonance. For the
simplest model of the resonance, that resulting from the circular
restricted three-body problem, the frequency for external resonances 
is given by
\be
\omega_0^2 = -3 j_1^2 {\cal C}_R n e^{|j_3|} \qquad {\rm where} 
\qquad {\cal C}_R = \left( {m_P \over M_\ast} \right) 
n \left[ f_d (\alpha) + \alpha^{-1} f_i (\alpha) \right] \, , 
\ee
where $n$ is the mean motion of the outer planet, $\alpha$ =
$a_1/a_2$, and the functions $f_d (\alpha)$ and $f_i (\alpha)$ are
given by the Laplace coefficients (see \S 8.5 of MD99). For odd order
resonances, the function $f_d < 0$, so that the corresponding
frequencies are real.  For even order resonances, $f_d > 0$, but the
equilibrium angle is shifted by $\pi$, and the frequencies are again
real (see MD99 for further discussion).  Notice also that $f_i$ is
nonzero only for the 2:1 resonance.  The integers $j_1$ and $j_3$
depend on the type of resonance. Although both integers are negative
for the cases of interest, the libration timescale only depends on the
absolute value. More specifically, the integer pair $(|j_1|, |j_3|)$
takes on the values (1,1), (3,2), and (2,1) for the 2:1, 5:3, and 3:2
resonances, respectively. For the values $e$ = 0.10 and $\mu = m_P /
M_\ast = 10^{-3}$, as used in the numerical simulations, we find that
$\omega_0^2/(3 \mu e n^2) \approx$ 2.0, 3.0, and 8.1 for the three
resonances. For these parameter values, the {\it square} of the
frequencies are spaced by factors of $\sim2$. This simple analytic
result is thus in qualitative, but not quantitative, agreement with
the numerical results. One should keep in mind that the orbital
elements that enter into these formulae (e.g., $e$ and $n$) vary over
the course of the simulations, so that comparisons are complicated.

If the migration rate ${\dot a}/a$ were the only relevant variable,
then one would expect that capture into resonance would be compromised
at a fixed value of the dimensionless parameter $a \omega_0$ / 
$|{\dot a}| = \omega_0 \atime$. The case that most closely meets this
expectation is that of migration with no eccentricity damping and no
turbulent forcing (shown in the lower right panel of Figure
\ref{fig:basicrat}). For this class of simulations, systems tend to
enter the 3:2 resonance states for higher migration rates than for the
2:1 resonances, where this trend is predicted (qualitatively) by the
simple theory outlined above. However, the fraction of systems in 5:3
resonance is not larger than the fraction in 2:1 resonance at large
migration rates, in spite of the 5:3 having a shorter libration period.
The 5:3 resonance is generally weaker, in the sense of being easier to
disrupt, than the first order resonances, and does not survive for
large migration rates. In terms of survival of the resonances, shown
by the solid curves, the fraction in 2:1 is generally larger for all
migration rates due to its greater stability.

For the case of migration with large eccentricity damping rates (see
the lower right panel of Figure \ref{fig:edamp}), the probability of
resonance survival shows the expected qualitative behavior: Each of
the resonances dominates (has the largest fraction) for a well-defined
range of migration rates.  The 2:1 resonance is by far the most
important for migration timescales longer than about $10^4$ yr. For
shoter timescales, there is a narrow window of migration rates where
the fraction of systems in 5:3 resonance shows a peak, and then the
3:2 resonance dominates for faster migration rates.  Although the
ordering of these results is consistent with theoretical expectations,
the maximum migration rates are spaced at larger intervals than the
factors of $\sqrt{2}$ suggested by the above analysis. Here, the large
eccentricity damping rates significantly change the dynamics and hence
the numerical values. Nonetheless, the qualitative trend holds up.

\begin{figure} 
\figurenum{6} 
{\centerline{\epsscale{0.90} \plotone{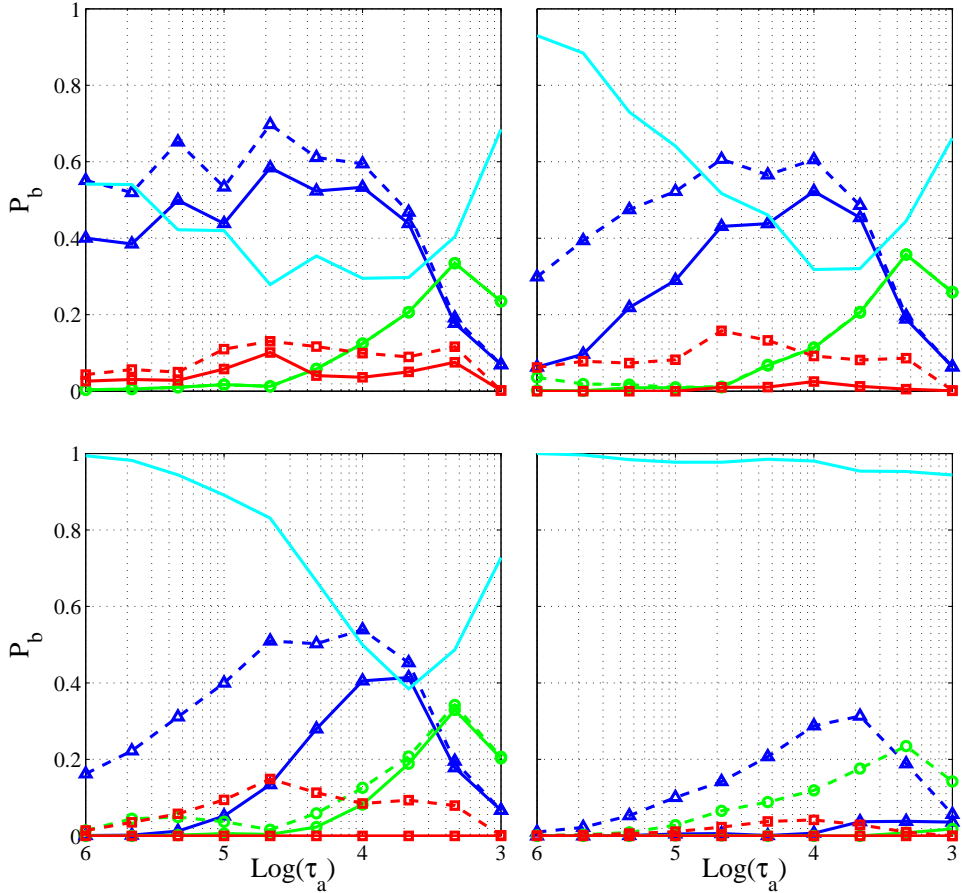} } } 
\figcaption{Effects of increasing turbulence on the survival of mean 
motion resonance. This two-planet system has planetary masses of 
$m_1 = 1 \mjup$ and $m_2 = 10 \mearth$. The four panels show the
survival fractions as a function of migration rate for increasing
levels of turbulence, as specified by the forcing strength 
$[(\Delta J)/J]_k$ per independent realization of the turbulent 
fluctuations. Results are shown for $[(\Delta J)/J]_k$ = 0 
(no turbulence), $10^{-4}$, $3 \times 10^{-4}$, and $10^{-3}$, 
where turbulence increases from upper left to lower right. In each panel,
the curves correspond to various resonances, including the blue curves
marked by triangles (2:1), red curves marked by squares (5:3), and
green curves marked by circles (3:2). The unmarked cyan curves show
the fraction of system that are not found in any of the mean motion
resonances. The solid curves show the fraction of systems in resonance
at the end of the migration epoch; the dashed curves show the largest
value of the fractions during the migration epoch.  The eccentricity
damping parameter $K$ = 1 for these simulations. } 
\label{fig:turbamp} 
\end{figure}

\begin{figure} 
\figurenum{7} 
{\centerline{\epsscale{0.90} \plotone{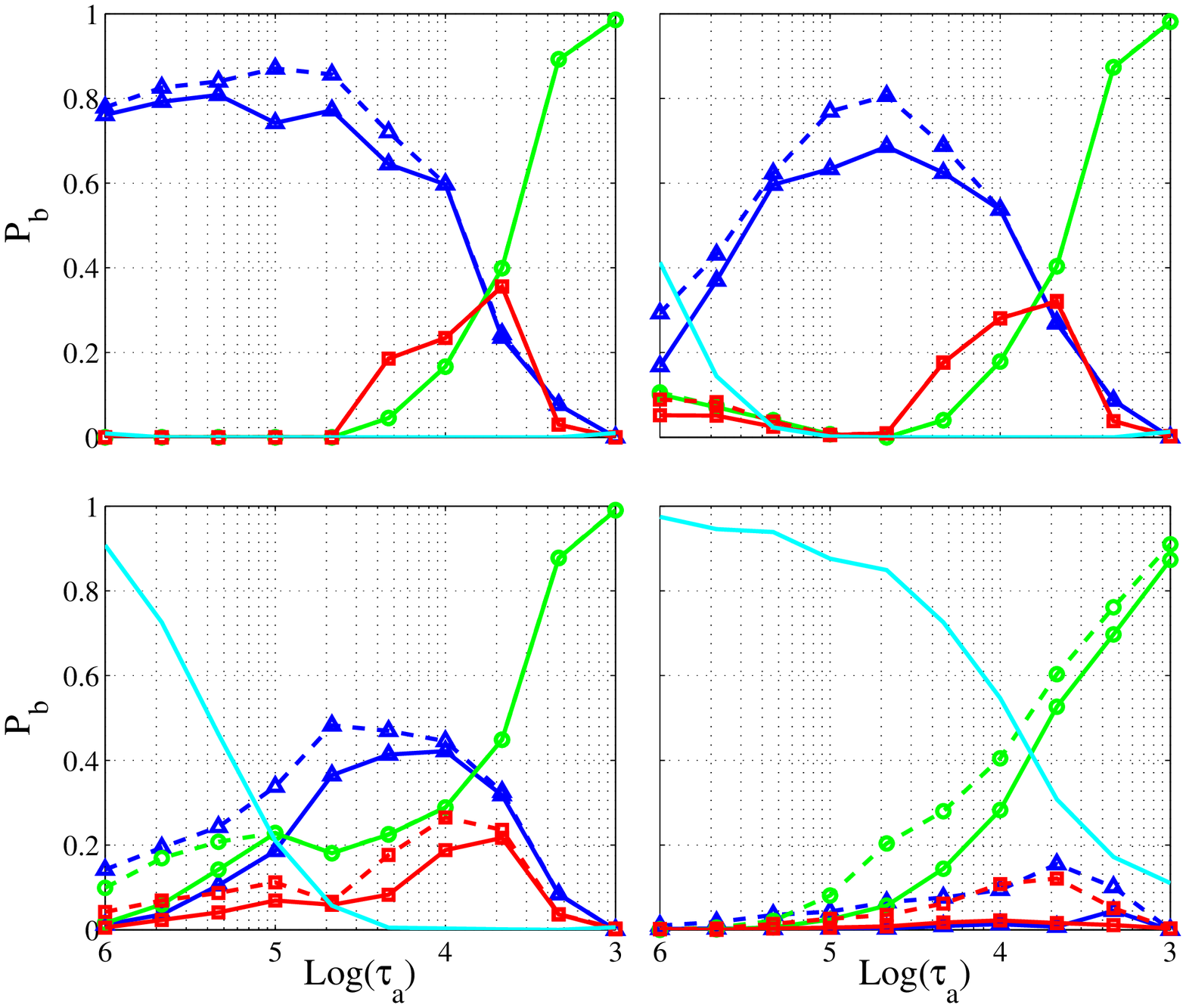} } } 
\figcaption{Effects of increasing turbulence on the survival of mean 
motion resonance for systems with eccentricity damping parameter $K$ =
10. Other system properties are the same as in Figure \ref{fig:turbamp}.
This system has planet masses $m_1 = 1 \mjup$ and $m_2 = 10 \mearth$. 
The four panels show the survival fractions as a function of migration
rate for increasing levels of turbulence.  Results are shown for
forcing strength $[(\Delta J)/J]_k$ = 0, $10^{-4}$, $3 \times
10^{-4}$, and $10^{-3}$, where turbulence increases from upper left to
lower right. In each panel, the curves correspond to various
resonances, including the blue curves marked by triangles (2:1), red
curves marked by squares (5:3), and green curves marked by circles
(3:2). The unmarked cyan curves show the fraction of system that are
not found in any of the mean motion resonances. The solid curves show
the fraction of systems in resonance at the end of the migration
epoch; the dashed curves show the largest value of the fractions
during the migration epoch. }
\label{fig:turbampten} 
\end{figure}

As the levels of turbulence increase, systems have greater difficulty
maintaining mean motion resonance.  This trend is quantified by the
simulations depicted in Figures \ref{fig:turbamp} and
\ref{fig:turbampten}.  These numerical experiments are carried out
using the standard case of a Jovian planet on the inside and an inward
migrating ``super-earth'' with mass $m_2$ = 10 $\mearth$. The
eccentricity damping rate is set at the standard values of $K=1$
(Figure \ref{fig:turbamp}) and $K=10$ (Figure \ref{fig:turbampten}).
As the amplitude of the turbulent fluctuations increases (from upper
left to lower right in both Figures), the general trend is for the
fraction of systems in resonance to decline significantly. The 2:1
mean motion resonance, which is the strongest and the first to be
encountered, is compromised for sufficiently rapid migration rate. As
the level of turbulence increases, the migration rate at which systems
leave the 2:1 resonance becomes lower (the curves shift to the right
in the figures).  We also note that the destructive action of
turbulence is more pronounced for the solid curves, i.e., for the
fraction of systems that remain in resonance at the end of the
migration time. Finally, as expected, we find that more resonant
systems survive for larger rates of eccentricity damping (compare
Figures \ref{fig:turbamp} and \ref{fig:turbampten}).

With the initial conditions used herein, where the planets are started
outside the 2:1 resonance, the faster 5:3 and 3:2 resonances are not
affected as severely by the presence of turbulence.  These other
resonances only arise when the migration rate is rapid, so that the
migration timescale is short and turbulence has little time to act.
For the 5:3 and 3:2 resonances, the probability curves shown in Figure
\ref{fig:turbamp} decrease slowly with increasing turbulent amplitude.
As expected, the largest effect arises for the largest turbulent
amplitude $[(\Delta J)/J]_k$ = $10^{-3}$, where the probability of
remaining in any of the resonant states is extremely low at the end of
the migration epoch; the fraction of systems not bound into resonance
is marked by the solid cyan curve, which is close to unity for all
migration rates. Note that the 3:2 resonance lasts the longest in the
face of increasing turbulence.  This apparent resilience arises
because the 3:2 cases are only present for fast migration rates, the
regime where turbulence has less time to act (it is not due to the
increased durability of the resonance).

\subsubsection{Equal Mass Planets} 

\begin{figure} 
\figurenum{8} 
{\centerline{\epsscale{0.90} \plotone{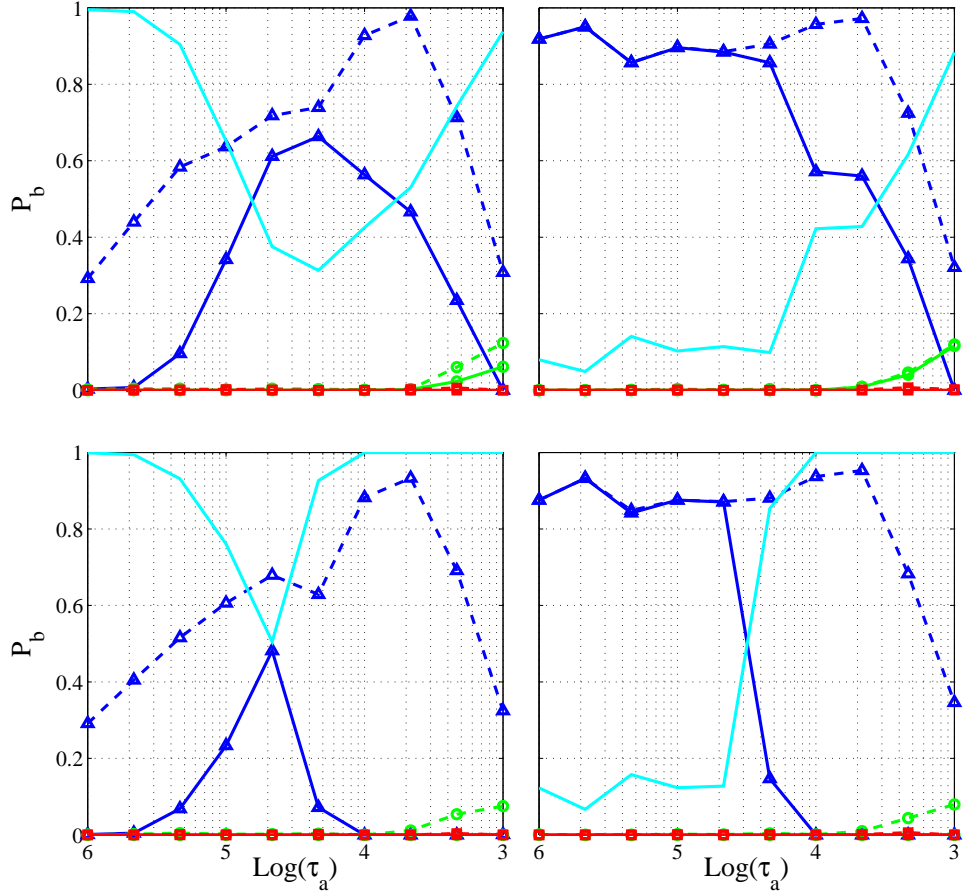} } } 
\figcaption{Effects of eccentricity damping and turbulent forcing on 
the survival of mean motion resonances for planetary systems containing two
Jovian planets ($m_1 = m_2 = 1 \mjup$).  The panels on the left
include turbulent forcing; the top panels include eccentricity damping
(where the eccentricity damping parameter $K$ = 1).  The lower right
panel shows the results with migration only. All of the panels show
the fraction of systems that remain bound in mean motion resonance as
a function of migration time scale (measured in yr). The curves
correspond to various resonances, including the blue curves marked by
triangles (2:1), red curves marked by squares (5:3), and green curves
marked by circles (3:2).  The unmarked cyan curve shows the fraction of
system that are not found in any of the mean motion resonances. Solid
curves show the fractions at the end of the migration epoch; dashed
curves show the peak values of the fractions during the migration
epoch. }
\label{fig:twojupiter} 
\end{figure}

Next we consider the case of two equal mass planets, with $m_1$ =
$m_2$ = $\mjup$. The results for survival of mean motion resonance are
shown in Figure \ref{fig:twojupiter}. The panels on the left include
the effects of turbulent forcing; the panels on the top include the
effects of eccentricity damping, where the parameter $K$ = 1 so that
the eccentricity damping timescale is the same as the migration
timescale. These results for two Jovian planets are significantly
different than those shown in Figure \ref{fig:basicrat} for the case
of a lower mass outer planet. One important effect of higher planetary
masses is to increase the levels of planet-planet interactions in the
systems. This effect, in turn, leads to greater libration widths for
systems that stay in resonance and a lower probability of remaining in
a resonant state.  As a result, the probability of the system residing
in either the 5:3 or the 3:2 resonance is significantly lower than in
the case of a less interactive system (compare Figures
\ref{fig:basicrat} and \ref{fig:twojupiter}). On the other hand, the
fraction of systems that remain in the 2:1 resonance is larger for the
more interactive (two jupiter) systems. 

\subsection{End States} 

During the course of the numerical integrations, the planetary systems
can end their evolution in a variety of ways. In many cases, the
systems remain bound together, even though mean motion resonance is
often compromised as described above. In many other cases, however,
planets can be lost through scattering encounters, through collisions
with each other, or via accretion onto the central star. This section
outlines the probabilities for each of these possible end states of
these dynamical systems.

\begin{figure} 
\figurenum{9} 
{\centerline{\epsscale{0.90} \plotone{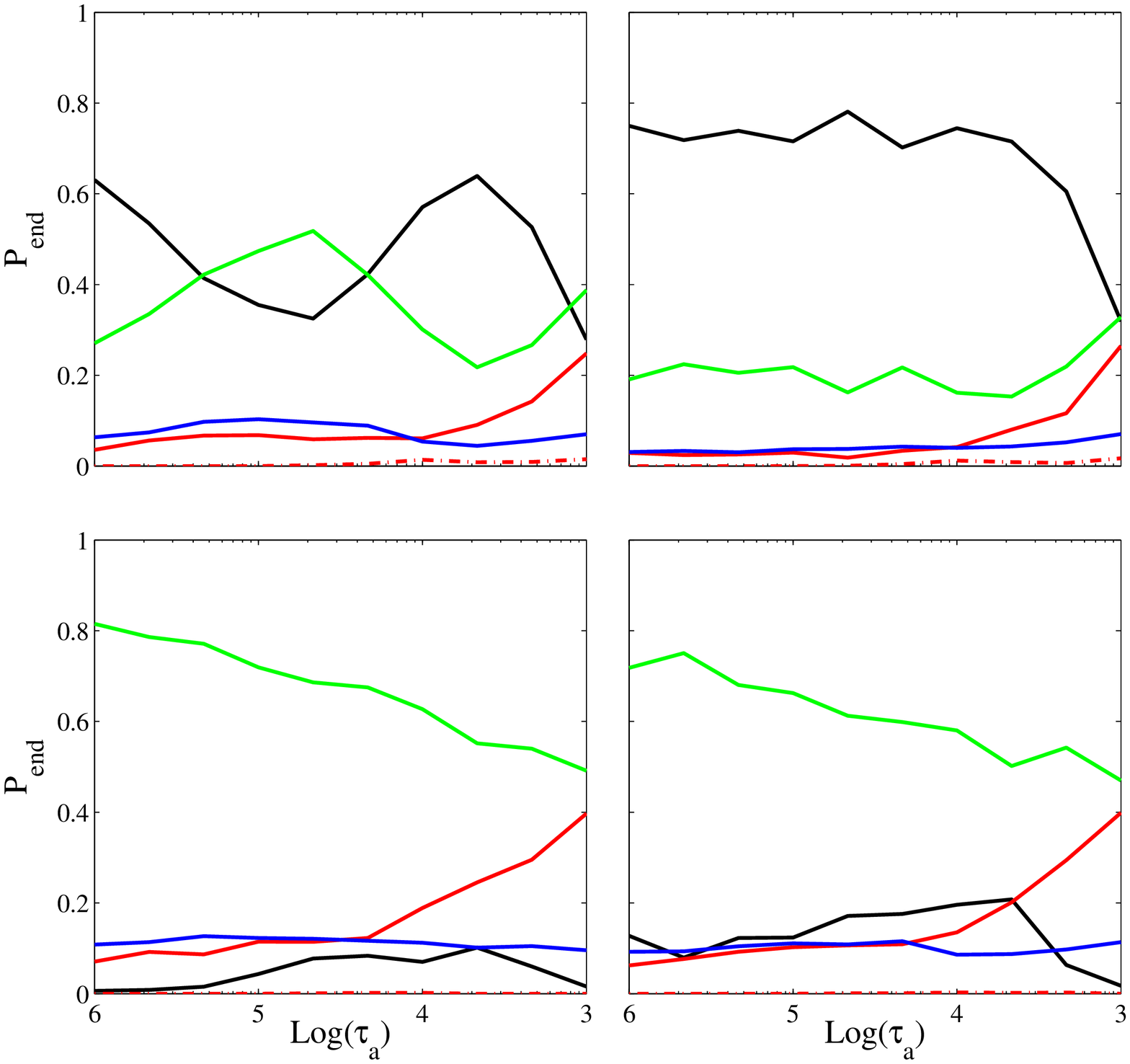} } } 
\figcaption{Probability of the planetary systems evolving into varying
end states for planet masses $m_1 = 1 \mjup$ and $m_2 = 10 \mearth$. 
Each panels shows the fraction of the systems that end their evolution
with a given end state, plotted here as a function of migration
rate. The end states represented here include survival of both planets
(black curves), planetary collisions (blue curves), ejection of a
planet (green curves), and accretion of a planet by the central star
(red curves). Four ensembles of simulations are depicted for migration
only (lower right panel), migration and eccentricity damping (with $K$
= 1; upper right), migration and turbulence (lower left), and all
three effects (upper left). }
\label{fig:endings} 
\end{figure}

For eccentricity damping parameter $K$ = 1, Figure \ref{fig:endings}
shows the likelihood of the planetary systems ending their evolution
in various possible end states for the standard case with inner planet
mass $m_1 = 1 \mjup$ and outer planet mass $m_2 = 10 \mearth$.  These
probabilities are shown as a function of migration rate for ensembles
of simulations with migration only, migration and eccentricity
damping, migration and turbulence, and for simulations including all
three effects. The evolution of these systems produces a wide variety
of outcomes, including survival of both planets for the entire
evolutionary time (shown by the black curves), ejection of a planet
(green curves), accretion by the central star (red curves), and
collisions between the planets (blue curves). As illustrated by the 
four panels in the figure, the corresponding probabilities depend
sensitively on the migration rates, eccentricity damping rates, and
the levels of turbulence.

Figure \ref{fig:endings} shows several trends. In general, the
probability for both planets to survive tends to decrease with
increasing values of the migration rate. This trend is expected,
because slow migration rates allow the systems to adjust as they
evolve; these cases with slow migration systematically exhibit less
overall action than cases with higher migration rates. One important
exception to this trend arises for the case of slow migration rates,
the inclusion of turbulent fluctuations, and no eccentricity damping
(see the lower left panel in Figure \ref{fig:endings}). In this
regime, migration time scales are long enough that turbulence has time
to act, which leads to loss of mean motion resonance (see the previous
section), a greater possibility of orbit crossing, and subsequent
planetary ejection. For this class of systems, the outer planet has
substantially less mass than the inner planet and is far more
susceptible to being lost. The outer planet is removed through
ejection, accretion onto the central star, and through collisions with
the inner Jovian planet. Note that the first two of these channels
dominate the third. 

Figure \ref{fig:endings} shows another trend: As the migration rate
increases, the probability of losing a planet through ejection
decreases, whereas the probability of losing a planet through
accretion onto the star increases.  One important physical property
that determines the relative number of accretion events versus
ejections is the location of the planet(s) in the gravitational
potential well of the star at the end of the migration epoch (when the
planets are likely to suffer close encounters).  The depth of the
stellar potential well at $a$ = 1 AU is approximately (30 km/s)$^2$,
whereas the depth of the potential well at the surface of Jupiter is
(43 km/s)$^2$. These scales are thus comparable.  For fast migration
rates, the outer planet is able to push the inner planet somewhat
farther inward, deeper into the stellar potential well, and hence the
probability of ejection decreases.

\begin{figure} 
\figurenum{10} 
{\centerline{\epsscale{0.90} \plotone{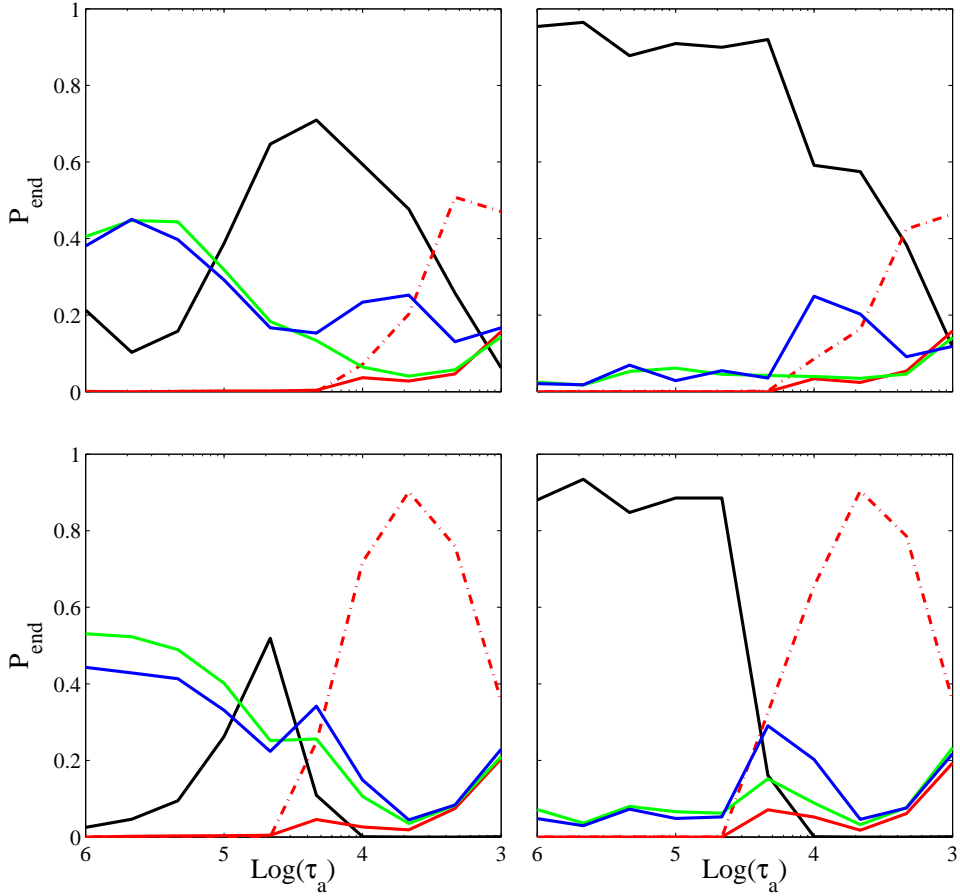} } } 
\figcaption{Probability of the planetary systems evolving into varying
end states for planet masses $m_1 = m_2 = 1 \mjup$.  Four ensembles of
simulations are depicted for cases with migration only (lower right
panel), migration and eccentricity damping (upper right), migration
and turbulence (lower left), and all three effects (upper left). The
end states represented here include survival of both planets (black
curves), planetary collisions (blue curves), ejection of a planet
(green curves), and accretion of a planet by the central star (red
curves). }
\label{fig:endings_jups} 
\end{figure}

Figure \ref{fig:endings_jups} shows the analogous plots for the channels
of planetary loss for systems initially containing two Jovian planets
($m_1 = m_2 = 1 \mjup$). The trends are roughly similar to the case
with lower mass outer planets: Planetary survival decreases with
increasing migration rate. Turbulence leads to planetary loss in the
regime of slow migration and no eccentricity damping, where the regime
of slow migration corresponds to migration time scales longer than
about $\atime$ = $3 \times 10^4$ yr. And, as the migration rate
increases, there is a shift from loss of planets through ejection to a
loss of planets through accretion onto the central star.  However, for
these systems with two Jovian planets, ejections, collisions, and
accretion events are on a more equal footing. One clear difference
from the case of low-mass outer planets is that planet-planet
collisions are more common (compare the blue curves in Figures
\ref{fig:endings} and \ref{fig:endings_jups}). The other significant
difference is that the inner planet is more often lost during
accretion events, rather than the outer planet (shown by the dotted
curves in Figure \ref{fig:endings_jups}).

For solar systems with sufficiently large values of the eccentricity
damping parameter $K$, most of the planets survive over the relatively
short timescales considered in this paper.  For example, for cases
with $K$ = 10, most systems remain intact and neither eject nor
accrete a planet. However, as shown by the comparison of Figures
\ref{fig:basicrat} and \ref{fig:basicten}, the fraction of systems
that remain in mean motion resonance for $K$ = 10 is only moderately
increased over the values obtained for $K$ = 1. The solar systems that
are not in resonance will often eject or accrete planets on longer
timescales, even in the absence of additional migration (e.g., Holman
\& Wiegert 1999; David et al. 2003). This issue should be addressed
with additional, longer term numerical integrations, but is beyond the
scope of this present work.

\section{MODEL EQUATIONS} 
\label{sec:model} 

In this section we derive a Hamiltonian model to describe the
migration of a pair of planets into mean motion resonance. In this
context, we want to find the simplest possible set of model equations
that captures the essential physics. Toward this end, we make a number
of simplifying assumptions. In particular, most of this discussion is
restricted to the case of a single resonance, which we take to be the
2:1 mean motion resonance; note that other resonances can be
considered in similar fashion.  In qualitative terms, this analysis
should apply to the variety of resonances that we consider in the
numerical simulations of Section \ref{sec:numerical}. The development
is parallel to previous treatments (Quillen 2006, Friedland 2001).

\subsection{Derivation}

As a starting point, we consider a test particle of mass $m_0$
orbiting in the same plane as a larger planet with mass $m_P$, which
is orbiting a star of mass $M_\ast$. The masses thus obey the ordering
\be
m_0 \ll m_P \ll M_\ast \, . 
\ee 
The orbital elements of the test particle are as follows: $\lambda$ is
the mean longitude, $M$ is the mean anomaly, $a$ is the semi-major
axis, $\varpi$ is the longitude of pericenter, and $e$ is the orbital
eccentricity. The analogous variables for the planet have the same
symbols but are denoted with the subscript `$P$' (see below).
The Poincar{\'e} coordinates (MD99) can be written 
\be
\lambda = M + \varpi \qquad {\rm and} \qquad 
\gamtwo = - \varpi \, , 
\ee 
with momentum variables of the form 
\be 
\ell = (G M_\ast a)^{1/2} \qquad {\rm and} \qquad 
\Gamma = (G M_\ast a)^{1/2} \left[ 1 - (1 - e^2)^{1/2} \right] \, . 
\label{defgamma} 
\ee 
The Hamiltonian can be written in the form 
\be
H = - {(G M_\ast)^2 \over 2 \ell^2} - \disturb \, , 
\ee
where $\disturb$ is the disturbing function due to the 
gravitational interaction between the test particle and the 
planet. 

Specializing to the case of a 2:1 mean motion resonance where the
planet is the inner body, we perform a canonical transformation
using the generating function
\be
F_2 = I (2 \lambda - \lambda_P) \, , 
\ee
which leads to the new variables 
\be 
I = - \ell/2 \qquad {\rm and} \qquad 
\psi = \lambda_P - 2 \lambda \, .
\ee 
The new Hamiltonian for the unperturbed problem, without the 
disturbing function, has the form 
\be
H_{0;new} = - {(GM_\ast)^2 \over 8 I^2} - I n_P \, , 
\ee
where $n_P$ is the mean motion of the planet. 

Next we express all quantities in dimensionless form and expand around
the resonance. Here, distances are measured in units of the semimajor
axis $a$, time is measured in units of $(a^2/GM_\ast)^{1/2}$, and mass
is measured in units of $M_\ast$.  If we define
\be 
\delta \equiv I - I_0 \qquad {\rm and} \qquad 
{1 \over 4 I_0^3} = n_P (t_0) \, , 
\ee 
the new Hamiltonian now reads 
\be 
K_{0;new} = {constant} - (n_P - 1) \delta - 
{3 \delta^2 \over 8 I_0^4} \, . 
\label{kform} 
\ee

We must now include the relevant terms from the disturbing function,
which provides an expansion in orders of eccentricity (of both the
test mass and the planet). Here we keep only the leading order term
(see MD99 and Quillen 2006) and write the Hamiltonian (from equation
[\ref{kform}]) in the form 
\be
K (\delta, \psi, \Gamma, \gamtwo) 
= - 6 \alpha^2 \delta^2 - (n_P - 1) \delta - 
2 \mu f_2 \alpha^{1/2} \Gamma + A \Gamma^{1/2} \cos (\psi - \varpi) \, ,
\ee
where $A$ is the expansion coefficient in the disturbing function and 
where we have used the fact that $I_0$ = $\alpha^{-1/2}/2$ for these 
units and choice of resonance. 

Following Quillen (2006), we perform another canonical transformation 
using the generating function 
\be
F_2 = J_1 (\psi - \varpi) + J_2 \psi \, ,
\ee
which leads to the new variables 
\be 
J_1 + J_2 = \delta \, , \qquad \phi = \psi - \varpi \, , \qquad 
J_1 = \Gamma \, , \qquad {\rm and} \qquad 
\theta = \psi \, , 
\ee 
and hence the new Hamiltonian 
\be
H = - 6 \alpha^2 \left(\Gamma^2 + J_2^2 \right) - 
\left[ 12 \alpha^2 J_2 + (n_P - 1) + 
2 \mu f_2 \alpha^{1/2} \right] \Gamma 
- (n_P - 1) J_2 + A \Gamma^{1/2} \cos \phi \, .
\ee
Since $J_2$ is conserved and constant terms can be dropped, 
the Hamiltonian can be simplified to the form 
\be 
H = 6 \alpha^2 \Gamma^2 + \left[ 12 \alpha^2 J_2 + (n_P - 1) 
+ 2 \mu f_2 \alpha^{1/2} \right] \Gamma 
- A \Gamma^{1/2} \cos \phi \, .
\ee
Next we rescale the momentum variable $\Gamma$ according to 
the transformation 
\be
\Gamma \to \left[ {6 \alpha^2 \over A} \right]^{2/3} \Gamma \, ,
\ee
and rescale the time variable so that the 
Hamiltonian $H$ is given by 
\be
H = \Gamma^2 + b \Gamma - \Gamma^{1/2} \cos \phi \, .
\label{hamworking} 
\ee
The parameter $b$ is thus given by 
\be
b = \left[ 12 \alpha^2 J_2 + (n_P - 1) 
+ 2 \mu f_2 \alpha^{1/2} \right] 6^{-1/3} (\alpha A)^{-2/3} \, . 
\ee
The first and third terms in square brackets are generally small
compared to unity. The central term vanishes on resonance, by
definition, but can be of order unity when the system is far from
resonance. As a result, the parameter $b$ provides a measure of how
far the system resides from a resonant condition. For this paper, 
we let the parameter $b$ evolve linearly with time so that the 
systems approach resonance ($b$ = 0) at a well-defined rate. 

Using the Hamiltonian with the form given by equation
(\ref{hamworking}), the equations of motion become 
\be
{d \Gamma \over dt} = - \Gamma^{1/2} \sin \phi \, , 
\ee 
and 
\be
{d \phi \over dt} = 2 \Gamma + b - {1 \over 2 \Gamma^{1/2}} 
\cos \phi \, . 
\ee
It is useful to define the reduced momentum variable 
$p \equiv \Gamma^{1/2}$ so that the equations of motion 
simplify to the forms
\be
2 {dp \over dt} = - \sin \phi \, , 
\label{dpdt} 
\ee
and 
\be
{d\phi \over dt} = 2 p^2 + b - {1 \over 2p} \cos \phi \, . 
\label{dphidt} 
\ee
Although this ansatz simplifies the equations of motion, note that the
variables $(\phi,p)$ are no longer canonical. We also note that this
change of variables in convenient for calculating curves in phase
space to analyze the dynamics (this exercise is carried out in the
Appendix).

\begin{figure} 
\figurenum{11} 
{\centerline{\epsscale{0.90} \plotone{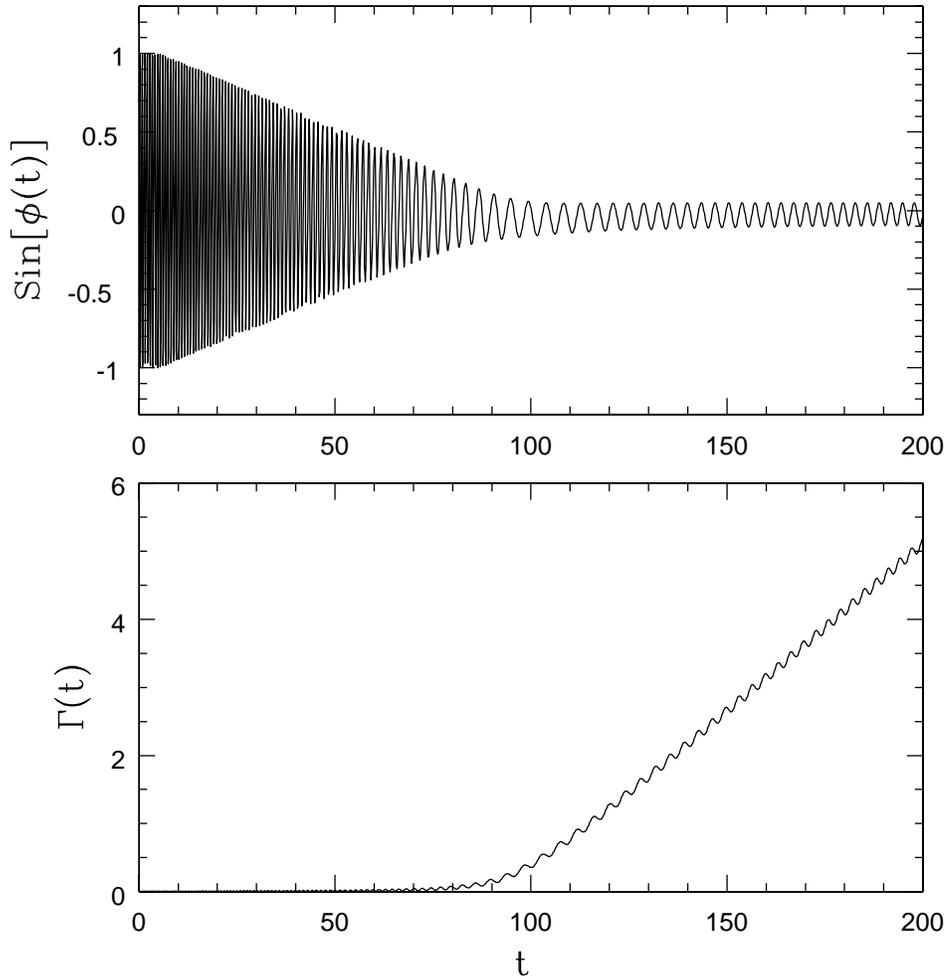} } } 
\figcaption{Time evolution of the resonance angle for a model 
system that becomes trapped in resonance. The top panel shows the
variable $\sin [ \phi(t)]$ versus time, for a starting value of
$\Gamma_0$ = 0.01 and a migration rate $db/dt$ = --0.1. The bottom
panel shows the time evolution of the momentum variable $\Gamma$. }
\label{fig:sineplot} 
\end{figure}

\subsection{Entry into Resonance} 

Using the model equations derived above, we can study the entry into
mean motion resonance as a function of the normalized migration rate
$db/dt$.  Here, the initial conditions are given by the starting
momentum $\Gamma_0$ and the starting value of the angle $\phi$.  We
choose fixed values of the momentum variable $\Gamma_0$ and then study
the probability of entering into resonance as a function of migration
rate $db/dt$. Since these systems often display extreme sensitivity to
their starting conditions, we must perform many realizations of the
numerical integrations for each pair $(\Gamma_0, db/dt)$, where each
realization uses a different value of the starting angle $\phi$. For
the sake of definiteness, we start the systems with $b = b_0$ = 10
(well outside of resonance) and let the resonance parameter evolve
according to the relation $b(t) = b_0 - (db/dt) t$. The systems thus
pass through resonance at time $t = b_0/|db/dt|$.

One example integration is shown in Figure \ref{fig:sineplot}, which
plots the quantity $\sin \phi$ (top panel) and the momentum variable
$\Gamma$ (bottom panel) as a function of time for a system that
becomes locked into mean motion resonance. In this case, the libration
width of the system steadily decreases with time until it reaches a
steady state near time $t$ = 100 (in dimensionless units). In this
case, $db/dt$ = 0.1, so that $t \sim 100$ corresponds to the time when
the system passes through resonance (as expected). The momentum
variable $\Gamma$ stays small until the system enters resonance, and
then grows steadily (see also the discussion of Quillen 2006).

\begin{figure} 
\figurenum{12} 
{\centerline{\epsscale{0.90} \plotone{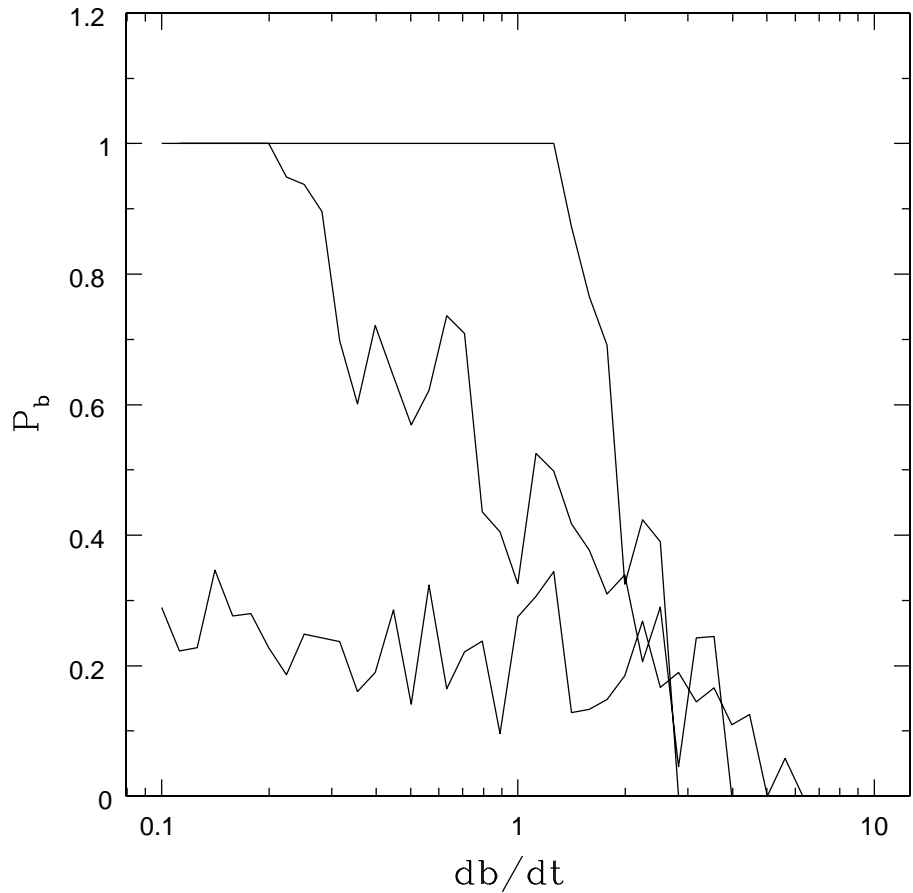} } } 
\figcaption{Fraction of systems that survive in mean motion resonance
as a function of migration rate $db/dt$. The three curves correspond 
to different initial conditions, where the the angular momentum variable 
$\Gamma_0$ = 0.1 (top curve), 1 (center curve), and 3 (bottom curve).
Each point on each curve shows the result of 1000 realizations of the
evolution, each with a randomly chosen starting angle. Note that this 
model system corresponds to the case of the 2:1 mean motion resonance. } 
\label{fig:survive} 
\end{figure}

As found previously (Quillen 2006), the probability of entering and
surviving in resonance decreases with increasing migration rate. This
trend is illustrated in Figure \ref{fig:survive}, which shows the
probability of achieving a resonant state versus the migration rate
$db/dt$. The three curves shown in the figure use different starting
values of the momentum variable $\Gamma_0$ = 0.1, 1, and 3.  Recall
that $\Gamma$ is related to the orbital eccentricity of the migrating
planet (equation [\ref{defgamma}]). Previous work shows that small
starting momentum generally leads to resonance capture, whereas larger
values generally do not (Quillen 2006); the value $\Gamma_0$ = 1
corresponds to the transition region.  For each value of the rate
$db/dt$, we have performed an ensemble of 1000 integrations, each with
a different starting value of the angular variable $\phi$. The
probability of capture decreases with increasing $db/dt$, but the
curves show a great deal of additional structure. The probability of
achieving resonance decreases near $db/dt$ = 1 and approaches zero for
somewhat larger values $db/dt \sim 3 - 5$.

Another clear trend is that increasing the initial value of the
momentum variable $\Gamma_0$ acts to decrease the probability of
entering into resonance. In other words, larger eccentricities tend to
compromise the chances of attaining resonance.  This finding is
consistent with the full numerical integrations of the previous
section, where eccentricity damping was found to allow for more
resonant states (see Figure \ref{fig:edamp}).

The leading order trend illustrated by Figure \ref{fig:survive} is
that resonant capture is more difficult with fast migration. This
result, obtained from the model equations of this section, is thus 
consistent with the results of the numerical simulations of Section
\ref{sec:numerical} (see Figures \ref{fig:criteria} -- 
\ref{fig:twojupiter}).  We can understand this effect through a simple
analysis: In the limit of large $db/dt = \gamma$, which we consider to
be a constant, the equation of motion for $\phi$ simplifies to the form
\be
{d \phi \over dt} = - \gamma t \qquad \Rightarrow \qquad 
\phi = - {1 \over 2} \gamma t^2 \, , 
\ee
where we have used the same sign convention as before.  The momentum
variable is then given by the remaining equation of motion, which can
be written in integral form
\be 
p - p_0 = 
{1 \over 2} \int \sin \left( {1 \over 2} \gamma t^2 \right) dt = 
{1 \over (2 \gamma)^{1/2} } \int \sin u^2 du = {1 \over 2} 
\left( {\pi \over \gamma} \right)^{1/2} S \left[ 
\left( {\gamma \over \pi} \right)^{1/2} t \right] \, , 
\label{fresnel}
\ee 
where $S(z)$ is the Fresnel integral (e.g., Abramowitz \& Stegun 1965).  
In the limit $z \to \infty$, $S(z) \to 1/2$, so the expression on the
right hand side of equation (\ref{fresnel}) approaches a constant
value $(\pi/16\gamma)^{1/2}$.  As result, the momentum variable $p$
approaches a constant, and hence does not grow, so the system does not
enter resonance. For a given starting value of the momentum variable
$\Gamma_0$, the critical value of the migration rate $\gamma$ can be
estimated by
\be
\gamma_c = {\pi \over 16 \Gamma_0} \qquad {\rm or} \qquad 
\gamma_c \, \Gamma_0 \approx 1/5 \, . 
\ee 
For comparison, in the set of simulations shown in Figure
\ref{fig:survive} with $\Gamma_0$ = 1, the probability of survival in
resonance $P_b$ falls below unity when the migration rate becomes
greater than $\gamma = db/dt \approx 0.2$; $P_b$ falls below 1/2 for
$\gamma > 1$ and goes to zero for larger values.

\begin{figure} 
\figurenum{13} 
{\centerline{\epsscale{0.90} \plotone{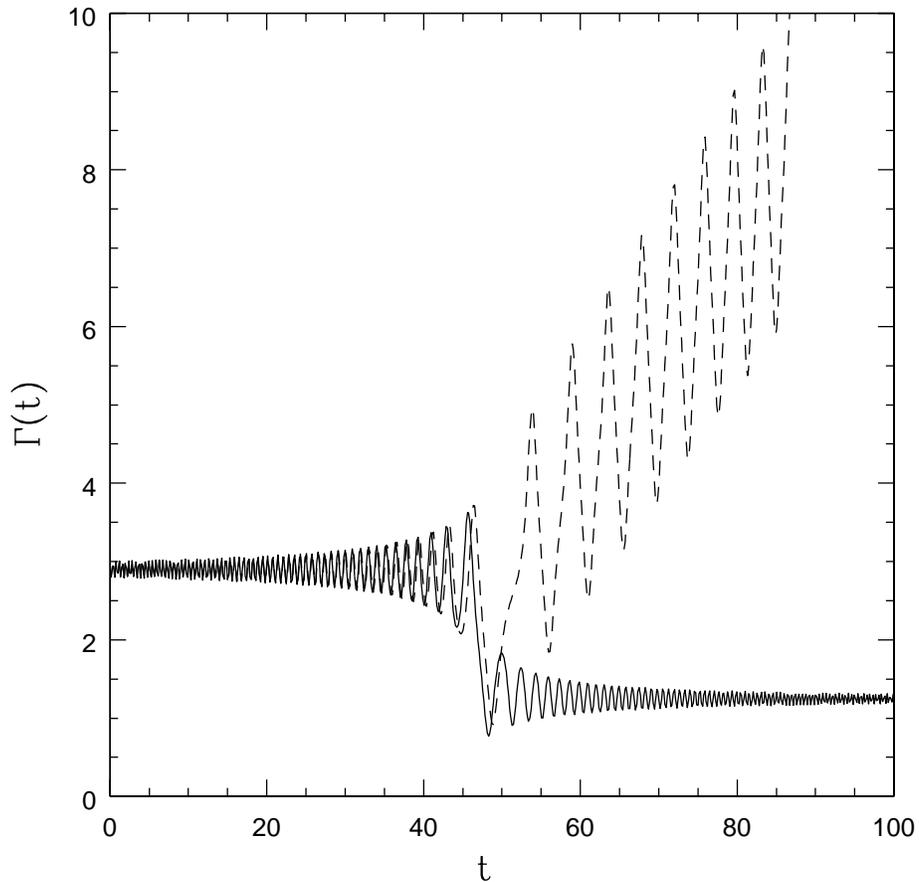} } } 
\figcaption{Comparison of the momentum evolution of two nearly 
identical systems. Both systems are started with the same values of
the phase space variables $(\Gamma_0, \phi_0)$. The migration rates
are taken to be $db/dt$ = 0.300 (solid curve) and $db/dt$ = 0.301
(dashed curve). This small difference in migration rate allows one 
system to enter into mean motion resonance (dashed curve), while 
the other continues to circulate (solid curve). }  
\label{fig:branch} 
\end{figure}

Another trend present in Figure \ref{fig:survive} is that small
variations in the migration rate can significantly change the
probability of resonant capture, especially for larger starting values
of the momentum variable.  The curves shown in the Figure display a
great deal of variation with $db/dt$; if the curves were plotted with
finer resolution in $db/dt$, the plot would show even greater
variation (and would not show resolved oscillations). This sensitivity 
to the migration rate can be illustrated further by plotting the time
evolution of two nearly identical systems, as shown in Figure
\ref{fig:branch}.  In this case, two systems are started with
$\Gamma_0$, the same angle $\phi_0$, and two different migration rates
$db/dt$ = 0.300 (solid curve) and $db/dt$ = 0.301 (dashed curve).  The
evolution of the two systems is nearly identical until about halfway
through the total time interval, when the second system becomes
locked into mean motion resonance (indicated by the growing values of
$\Gamma$), whereas the first system continues to circulate with its 
momentum variable exhibiting a decreasing amplitude. 

This effect can be (roughly) understood as follows: Suppose we
consider circulating solutions such that $\phi \approx \omega t$. 
The equation of motion for the angle $\phi$ then implies that
\be
\omega \sim 2 p^2 + b - {1 \over 2p} \cos\phi \, .  
\label{omega} 
\ee
The corresponding solution for the reduced momentum variable $p$ 
then becomes 
\be
p(t) \sim A + {1 \over 2 \omega} \cos \omega t \, , 
\label{psimple} 
\ee
where $A$ is a constant. In this context, the parameter $b$ starts at
a positive value (outside resonance) and then decreases.  The relation
(\ref{omega}) indicates that $\omega$ must decrease with time, so that
the amplitude of the oscillations of momentum increase with time as
the frequency decreases. Near the point where $\omega \to 0$, however,
the oscillation amplitudes are large and the frequency is small. The
system must then match onto one of the possible solutions for late
times when $b$ is large and negative. One solution corresponds to
$\omega \to b$ (see relation [\ref{omega}]); in this case, equation
(\ref{psimple}) indicates that the momentum variable will oscillate
with increasing frequency and decreasing amplitude (shown by the solid
curve in Figure \ref{fig:branch}). Although the momentum variable
oscillates, the resonance angle circulates for this case.  A second
solution exists for sufficiently large $p$; in this case, the equation
of motion (\ref{dphidt}) for the variable $\phi$ takes the approximate
form
\be
{d \phi \over dt} \approx 2 p^2 + b \, . 
\ee
This equation can be combined with the momentum equation (\ref{dpdt}) 
to obtain the result 
\be
{d^2 \phi \over dt^2} + 2 p \sin \phi + {db\over dt} = 0 \, , 
\ee
which is a type of pendulum equation, and hence allows for librating
solutions for the angle $\phi(t)$. This class of solution is depicted
by the dashed curve in Figure \ref{fig:branch}. 

\section{CONCLUSIONS} 
\label{sec:conclude} 

This paper studies the entry of planetary systems into mean motion
resonance, and the subsequent survival of resonant configurations,
with a focus on how the migration rate, eccentricity damping rate, and
turbulence levels affect the results. Our basic findings can be
summarized as follows:

In agreement with previous studies, we find that an inward migrating
planet naturally becomes locked into mean motion resonance when it
becomes sufficiently close to an inner planet. If the migration rate
is too fast, then mean motion resonance cannot be maintained. This
trend arises in both full numerical integrations of the 3-body system
with 18 phase space variables (Section \ref{sec:numerical}), and in
model equations (Section \ref{sec:model}), in agreement with previous
results (e.g., Quillen 2006). In rough terms, the probability of
staying in resonance is a decreasing function of the migration rate;
this probability (effectively) vanishes when the migration rate
exceeds the frequency of the resonant state.  As the migration rate
increases, the frequency of the resonances that the systems can
maintain also increases. For example, the three strongest resonances
considered here are the 2:1, 5:3, and 3:2, in increasing order of
frequency. As the migration rate increases, the systems become more
likely to pass through the 2:1 resonance and then become locked into
the 5:3. For even larger migration rates, the systems cannot maintain
5:3 resonance but enter into the 3:2 resonance. Figures
\ref{fig:basicrat} -- \ref{fig:twojupiter} all show this basic
trend. This general trend continues to hold up in the presence of
additional processes, such as eccentricity damping and turbulent
forcing; however, the critical values of the migration rate change, as
described below.

Eccentricity damping acts to maintain mean motion resonance (again, in
agreement with expectations; see Lecoanet et al. 2009). As a general
rule, larger eccentricity damping rates result in more systems
maintaining resonant configurations (see Figure \ref{fig:edamp}). For
a relatively non-interactive system (here we use $m_1 = 1 \mjup$ and
$m_2 = 10 \mearth$), a substantial increase in resonance survival is
realized with eccentricity damping parameter $K \ge 1$, where roughly
half the systems survive (Figure \ref{fig:basicrat}).  This survival
fraction increases to $P_b \sim 0.75$ for a larger eccentricity
damping parameter $K$ = 10 (Figure \ref{fig:basicten}).  In order to
increase the probability of survival close to unity (for relatively
``slow'' migration rates with $\atime > 3 \times 10^4$ yr), the
eccentricity damping parameter must be increased to about $K \ge
100$. This level of eccentricity damping can be realized in radiative
disk models (e.g., Bitsch \& Kley 2010).

This work also shows that turbulence acts to compromise mean motion 
resonance, in agreement with previous studies (Adams et al. 2008,
Lecoanet et al. 2009, Rein \& Papaloizou 2009). Because turbulence,
with the expected amplitudes, requires a long time to act, it
primarily affects those systems with slow migration rates.  We can
define an effective timescale for turbulent fluctuations to affect
resonances through the following heuristic argument.  For a stochastic
process, the system accumulates changes in angular momentum as a
random walk; after $N_S$ steps the angular momentum changes by
$N_S^{1/2} (\Delta J)_k$, where $(\Delta J)_k$ is the typical angular
momentum fluctuation per step. As an order of magnitude estimate, the
angular momentum of the resonant configuration is given by $J_{orb}
\omega_0 / \Omega$, where $\omega_0$ is the frequency of the resonance
and $J_{orb}$ is the orbital angular momentum.  The number of steps
required to compromise the resonance is then given by $N_S > [(\Delta
  J)_k/J_{orb}]^{-2} (\omega_0/\Omega)^2$.  The time required for an
independent realization of the turbulent fluctuations is approximately
the orbit time, so that the corresponding time scale becomes 
\be 
\ttime \approx {2 \pi \over \Omega} 
\left[ {(\Delta J)_k \over J_{orb}} \right]^{-2} 
\left( {\omega_0 \over \Omega} \right)^2 \, \approx 
3 \times 10^4 \, {\rm yr} \, 
\left[ {(\Delta J)_k / J_{orb} \over 10^{-4} } \right]^{-2} \, , 
\ee 
where the second equality scales the result to the parameters used in
this study. In order for turbulence to have a significant effect, the
migration timescale must be longer than this value.  As shown in
Figures \ref{fig:turbamp} and \ref{fig:turbampten}, turbulence
compromises mean motion resonance for slower migration rates, more
specifically for migration time scales $\tau_a = - a/{\dot a} > 10^5$
yr.

The above results can be summarized in terms of the four timescales in
this problem: the migration timescale $\atime$, the eccentricity
damping timescale $\etime$, the timescale $\ttime$ for turbulence to
act, and the libration timescale $\rtime$ of the mean motion
resonance. The relative ordering of these timescales determines much
of the dynamics. The numerical integrations (Section
\ref{sec:numerical}), the model equations (Section \ref{sec:model}),
and previous work (Quillen 2006) all show that planetary systems have
difficulty entering and maintaining mean motion resonance when $\atime
< \rtime$. Eccentricity damping allows more resonances to survive
provided that $\etime < \atime$ (see Figure \ref{fig:edamp} and
Lecoanet et al. 2008).  On the other hand, turbulence acts to destroy
resonances when $\ttime < \atime$ (see Figures \ref{fig:turbamp} and
\ref{fig:turbampten}, Adams et al. 2008, Rein \& Papaloizou 2009).

Although the trends outlined above are robust, the boundaries between
the various regimes are not sharp, and are subject to a number of
complications: First we note that the condition for passing through
resonance, $\atime < \rtime$, should be written in the more general
form $\atime < A \rtime$, where the factor $A$ depends on the details
of the system. For example, planetary systems with larger eccentricity
are generically less stable, so the factor $A$ will vary with orbital
eccentricity (e.g., see Figure \ref{fig:survive}).  Similarly, systems
with larger planetary masses are more interactive, so that the
parameter $A$ should increase with mass.  Each type of resonance has a
different libration timescale $\rtime$. In addition, the different
resonances have different strengths, as determined by the depth of the
effective potential well that the resonance angle resides within (and
this effect can be incorporated into the factor $A$ for a given
resonance). The libration timescale is also affected by the other
variables such as the migration timescale $\atime$ and/or the
eccentricity damping timescale $\etime$.  

One of the challenges facing applications of these ideas to extrasolar
planets is that many systems are expected to have comparable
timescales so that $\atime \sim \etime \sim \ttime$. All three of
these timescales are often longer than the typical libration time
$\rtime$, so that mean motion resonance is not usually compromised by
fast migration alone.  Instead, resonance configurations are
compromised by a combination of too rapid migration, too much
eccentricity excitation (not enough damping), and turbulent forcing
acting over long spans of time.  We also stress that these systems
display sensitive dependence on their initial conditions (e.g., Figure
\ref{fig:branch}), so that systems in essentially the same regime of
parameter space can result in widely different outcomes. These
differences are important, because migrating planets that maintain
resonance stand a much greater chance of survival (see Figures
\ref{fig:endings} and \ref{fig:endings_jups}).

Finally we note that planetary systems will continue to evolve after
the removal of disk material from the system. When the gaseous disk is
gone, the forcing terms that lead to migration, eccentricity damping,
and turbulent forcing will vanish. However, the system will continue
to evolve through gravitational forces.  Planetary systems that are
deep in mean motion resonance are expected to survive over long spans
of time; on the other hand, systems that are near --- but not in ---
resonance will often be disrupted over these longer time scales (e.g.,
Holman \& Wiegert 1999; David et al. 2003). 

\acknowledgements 

This work was supported in part by the Michigan Center for Theoretical
Physics. FCA is supported by NASA through the Origins of Solar Systems
Program via grant NNX07AP17G.  AMB is supported by the NSF through
grant DMS-0907949.  In addition, AMB and FCA are jointly supported by
Grant Number DMS-0806756 from the NSF Division of Applied Mathematics.
Portions of this work were carried out at the Kavli Institute for
Theoretical Physics, U. C. Santa Barbara, and was supported in part by
the National Science Foundation under Grant No. PHY05-51164.

\appendix 
\section{APPENDIX: PHASE SPACE ANALYSIS} 
\label{sec:phase} 

This Appendix discusses the phase plane for the model equations
developed in Section \ref{sec:model}. This analysis determines the
number of allowed regions in phase space, and hence places constraints
on the allowed dynamics.  Given the equations of motion (\ref{dpdt})
and (\ref{dphidt}), the basic equation for curves in phase space has
the form
\be
{dp \over d \phi} = - {\sin \phi \over 
4 p^2 + 2b - (\cos \phi) / p} \, . 
\label{phaseone} 
\ee
If we consider the parameter $b$ to be fixed, this equation can be
integrated directly to find an implicit solution of the form
\be
p \cos \phi = \left( p^4 - p_0^4 \right) + 
b \left( p^2 - p_0^2 \right) + p_0 \, , 
\ee
where $p_0 = p (\phi=0)$, by definition. This equation can be written 
in the alternate form 
\be
p^4 + b p^2 - p \cos \phi = E \qquad {\rm where} \qquad 
E \equiv p_0^4 + b p_0^2 - p_0 = {constant} \, . 
\label{solution} 
\ee

\subsection{Limiting Forms} 

In the limit of large $p \gg 1$, we can ignore the cosine term 
in the denominator of equation (\ref{phaseone}) and find the 
approximate solution 
\be
{4 \over 3} \left( p^3 - p_0^3 \right) + 2 b (p - p_0) = 
\cos\phi - 1 \, . 
\ee
This result can be rewritten in the form 
\be
- \sin^2 (\phi/2) = (p - p_0) \left\{ {2 \over 3} 
\left[ p^2 + p p_0 + p_0^2 \right] + b \right\} \approx 
(p - p_0) \left\{ 2 p_0^2 + b \right\} \, . 
\ee
Note that in the limit of large $p$, $|p-p_0| \ll p$, and 
the two expressions in the above equation are the same to 
leading order in the parameter $|p-p_0|/p$. 

In the limit of small $p \ll 1$, we can ignore the $p^2$ term
in equation (\ref{phaseone}) and find the solution 
\be
p \cos \phi = b \left( p^2 - p_0^2 \right) + p_0 \, . 
\ee
For sufficiently small $p$, this expression reduces to the simpler
form 
\be
p \cos \phi = p_0 \, . 
\ee 

\subsection{Regimes of Solutions} 

The solution for the phase space curves, given by equation
(\ref{solution}), can allow for multiple roots. We first note that the
parameters $b$, $E$, and $\cos\phi$ can all be both positive and
negative. As a result, the number of roots for $p$ will vary.

Case I: $b > 0$, $E > 0$: In this case, only one root for $p$ exists
for all values of the angle $\phi$ (or $\cos\phi$). For small values
of the energy $E$, the solutions for $p$ get small for negative values
of $\cos\phi$.

Case II: $b < 0$, $E > 0$: For $\cos\phi > 0$, only one solution for
$p$ exists.  For $\cos\phi < 0$, however, there can be multiple roots
provided that $|b|$ is large enough and the energy $E$ is small
enough. The conditions required for multiple roots is that 
\be 
|b| > {8 \over 27} \cos^2\phi \qquad {\rm and} 
\qquad 4 |b| E < \cos^2 \phi \, .  
\ee 
In the regime where the ``extra'' roots arise, $p \ll 1$, and the
solutions reduce to the approximate form 
\be 
p = {\cos \phi \pm \left[ \cos^2 \phi - 4 |b| E \right]^{1/2} 
\over 2 |b|} \, .  
\ee

Case III: $b > 0$, $E < 0$: There are no solutions for $\cos\phi < 0$.
For the case $\cos \phi > 0$, there are either two solutions for small
$E$, or no solutions.

Case IV: $b < 0$, $E < 0$: For $\cos\phi > 0$, there are two solutions
for small $E$ and no solutions if $E$ is too large and negative. For
the case $\cos\phi < 0$, the two solutions disappear if $|b|$ is too
small, where the critical value (for the limiting case $\cos\phi = -1$) 
is given by
\be
|b|_c = {3 \over 4^{1/3}} \approx 1.587 \, . 
\ee
For cases of interest, the parameter $b$ becomes large and negative. 
For bound states, the energy also becomes large and negative. In this
regime, the phase curves become almost independent of angle $\phi$, with 
\be
p^2 \approx {|b| \pm \left[ b^2 + 4 E \right]^{1/2} \over 2} \, . 
\ee

\begin{figure} 
\figurenum{14} 
{\centerline{\epsscale{0.90} \plotone{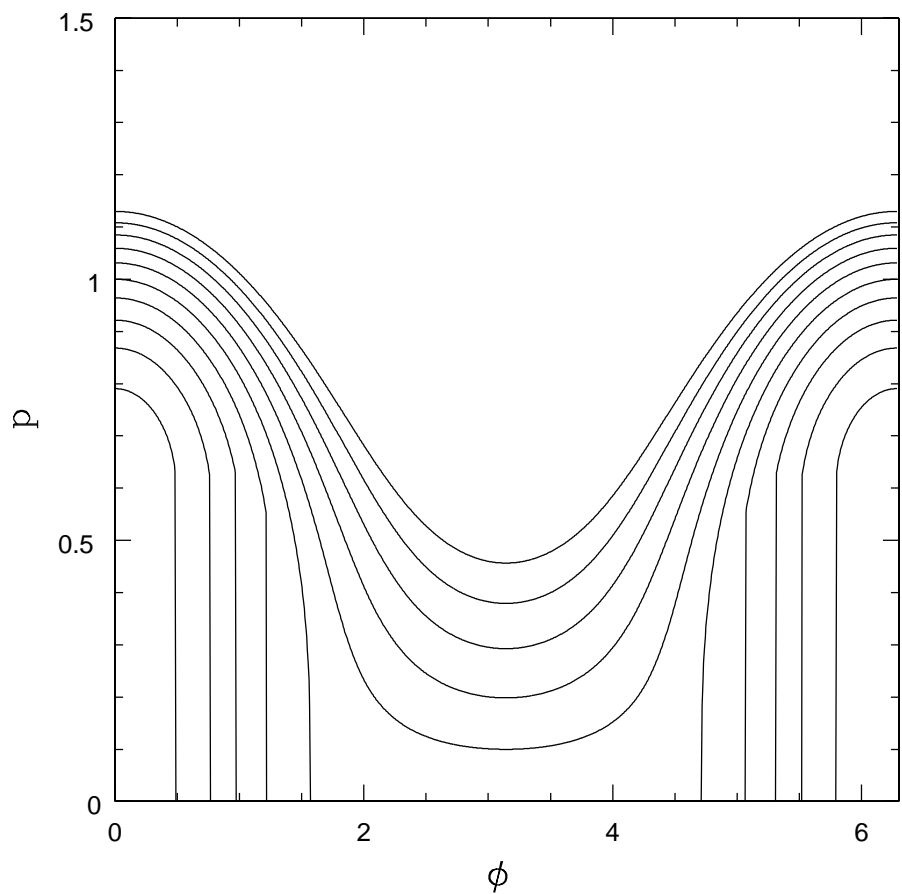} } } 
\figcaption{Phase plot for the case $b = 0$ where systems can enter
resonance. The various curves show decreasing values of energy from 
$E$ = 0.5 (top) down to $E = - 0.4$ (bottom). Note that as $E$ falls 
below zero, solutions for $p$ do not exist for negative values of 
$\cos\phi$. }
\label{fig:phase} 
\end{figure}

Figure \ref{fig:phase} shows one sample phase plot for the case where
$b$ = 0, which corresponds to systems that are passing through the
resonant condition.  For this case, as the energy variable $E$
decreases from positive to negative values, the phase curves change
their shape: For positive $E$, solutions for the momentum variable $p$
exist for all values of the angle $\phi$; for negative $E$, solutions
for $p$ exist for a limited range of angles. These isolated regions,
which become narrower as the energy $E$ grows more negative,
correspond to oscillatory solutions in $\phi$ (such as that shown in
Figure \ref{fig:sineplot}).

\end{document}